\documentclass{aa}  
\usepackage{graphicx}
\usepackage{txfonts}
\begin{document}
   \title{3.2\,mm lightcurve observations of (4)~Vesta and (9)~Metis with the
          Australia Telescope Compact Array}

   \titlerunning{3.2\,mm lightcurve observations}
   \author{T.\ G.\ M\"uller
          \inst{1}
          \and
          P.\ J.\ Barnes
	  \inst{2}
          }

   \offprints{T.\ G.\ M\"uller}

   \institute{Max-Planck-Institut f\"ur extraterrestrische Physik,
              Giessenbachstra{\ss}e, D-85748 Garching, Germany,\\
              \email{tmueller@mpe.mpg.de}
	      \and
	      School of Physics A28, University of Sydney,
              Sydney NSW 2006, Australia,\\
              \email{peterb@physics.usyd.edu.au};
             }

   \date{Received TBD; accepted TBD}

% \abstract{}{}{}{}{} 
% 5 {} token are mandatory
 
  \abstract
   {\object{(4)~Vesta} and \object{(9)~Metis} are large main-belt asteroids with high albedos.
    There are strong indications for heterogeneous surfaces for both targets
    from imaging techniques in the visible and near-IR range, very likely connected
    to impact structures. Despite that, the thermal spectral energy distributions
    from mid-IR to the mm-range have, until now, been consistent with a homogeneous regolith-covered surface
    and the thermal light-curves are dominated by the shape and spin vector
    properties.}
   {With millimetre-observations at 93.0 and 95.5\,GHz we tried to characterise the emission properties
   of the surface material. The coverage of the full rotation period allowed
   a detailed study of the heterogeneity of the surface. 
   }
   {We combined our carefully-calibrated mm-observations with sophisticated
    thermophysical modelling techniques. In this way it was possible to derive
    emissivity properties and to disentangle the effects caused by shape,
    albedo or various thermal properties.}
   {The rotationally averaged fluxes are explained very well
   by our thermophysical model techniques when using an emissivity in the
   mm-range of about 0.6 for (4)~Vesta and about 0.7 for (9)~Metis.
   The mm-lightcurves follow for a large fraction of the
   rotation period the shape-introduced variations. The rotational
   phases with clear deviations are connected to structures
   which are visible in the HST images of (4)~Vesta and the Keck AO-images
   of (9)~Metis. The observed lightcurve
   amplitudes are peak-to-peak $\sim$30\% for (4)~Vesta and $\sim$25\% for
   (9)~Metis, while the shape-related amplitudes are only 5 and 4\%, respectively.}
   {The emissivities at mm-wavelengths are lower than in the far-IR, confirming
   that particles with sizes of about 100\,$\mu$m influence the mm-behaviour. Previously
   identified bright spots at visible/near-IR wavelength are connected to sharp
   emissivity drops. The dark Olbers region on (4)~Vesta
   causes an excess in mm-emission on top of the shape introduced light-curve.
   The thermophysical model predictions match the overall
   flux levels very well, but cannot reproduce certain lightcurve features
   due to the lack of information on the grain size distribution.
   The 3-mm observations are very powerful for the study of surface
   heterogeneities.}


   \maketitle
%
%________________________________________________________________

\section{Introduction}

(4)~Vesta and (9)~Metis are large main-belt asteroids with
well-characterised shape, spin-vector, size and albedo properties
(e.g., Thomas et al. \cite{thomas97}; Torppa et al. \cite{torppa03};
Storrs et al. \cite{storrs99}; Marchis et al. \cite{marchis06}).
Both high-albedo targets also have indications of albedo variations on their surfaces
(Binzel et al. \cite{binzel97};
Nakayama et al. \cite{nakayama00};
Marchis et al. \cite{marchis06}).

(4)~Vesta is a V-type asteroid and it is believed to be the
parent body of the Vestoids (Binzel \& Xu \cite{binzel93}).
Radioisotope chronology from the howardite, eucrite, and diogenite
(HED) meteorites is correlated with a (4)~Vesta origin.
Most importantly (4)~Vesta has experienced significant excavating events,
most notably indicated by the huge crater near its southern pole
(Thomas et al. \cite{thomas97}). It is the first of the two asteroid
targets to be visited by DAWN (http://dawn.jpl.nasa.gov/;
Vernazza et al. \cite{vernazza05}). (4)~Vesta's visual lightcurve
is dominated by the influence of the albedo variations and
standard lightcurve inversion techniques failed to produce
a reliable shape model (Kaasalainen, priv. comm.), but high
resolution HST imaging allowed a solution for the shape and spin
vector (Thomas et al. \cite{thomas97}).

(9)~Metis is an S-type asteroid, which indicates a silicate and metal rich
composition, mainly olivines, pyroxenes and metals in various percentages.
Spectrophotometric similarity with \object{(113)~Amalthea} revealed a probable
compositional link between these two objects. A plausible common
parent body was estimated to have been between approximately 300
and 600 km in diameter (Kelly \& Gaffey \cite{kelly00}), but the search
for nearby companions and family members was not successful (Storrs et al.\cite{storrs05}).
Sophisticated lightcurve inversion techniques allowed a solution for a
shape and spin vector which fits nicely the available
visual lightcurves (Torppa et al. \cite{torppa03}). The shape and
spin-vector solution is also in excellent agreement with adaptive
optics images presented by Marchis et al. (\cite{marchis06}). However,
those authors noticed very strong intensity 
variations on the surface, but attributed them mainly to changing 
surface scattering properties.

Both targets were extensively observed at various wavelengths during
the last decades. For (4)~Vesta there are also a large number of thermal
observations from IRAS (Tedesco et al. \cite{tedesco02}),
ISO (M\"uller \& Lagerros \cite{mueller98}, abbreviated as M\&L 1998 in
following) and also at sub-millimetre and
millimetre wavelength (e.g., Redman et al. \cite{redman98}), while
for (9)~Metis the infrared observations are limited to very few ISO
(M\&L \cite{mueller98}) and MSX observations
(Tedesco et al. \cite{tedesco02a}). Due to their size and proximity
both main-belt asteroids are bright at millimetre-wavelengths and
easy to observe.

Our main objective was to obtain high-quality
measurements of the 3\,mm flux density of large main-belt asteroids
over their full rotation period. Since the mm-lightcurves are
not very much affected by albedo variations on the surface, we
wanted to see and confirm if the shape model can explain possible
brightness variations. As a second goal we were interested in the
%testing and calibrating our thermophysical model with respect
mm-emissivities of such bodies. Redman et al. (\cite{redman92})
reported a very low submm/mm-emissivity for (4)~Vesta and interpreted
their findings with the presence of a dusty, porous regolith.
How does the emissivity compare at 3\,mm? Furthermore, is the emissivity
of (9)~Metis, which has also a high albedo, comparable to that of (4)~Vesta?

Johnston et al.\ (\cite{johnston89}) performed microwave observations
of \object{(2)~Pallas}, (4)~Vesta and \object{(10)~Hygiea}. They 
concluded that the behaviour of these targets at cm-wavelengths
is dominated by a layer of material with the physical properties
of finely divided dust. They proposed to use mm-observations to
identify those areas with different regolith properties, reflected
in a rotational phase dependence of the brightness temperatures.
We present now for the first time such rotationally resolved
mm-observations. The data analysis and calibration are given
in Sect.~\ref{sec:obs_cal} and the observational results in
Sect.~\ref{sec:obs_res}.
We applied a thermophysical model to our observations and derive the important
thermophysical quantities. The methods used are described in
Sect.~\ref{sec:tpm}. In Sect.~\ref{sec:discussion} we combine the
TPM predictions and the observations and discuss the possible implications.

%__________________________________________________________________

\section{Observation and calibration}
\label{sec:obs_cal}

% In the time available we determined we could accomplish this goal
% for only these two asteroids; further tests with other asteroids will
% require more observations.
% In addition, because of the short sample times for the data acquisition,
% we could generate time-resolved 3\,mm-lightcurves during each
% body's sidereal rotation period, each of which is about 5\,h.

Observations of (4)~Vesta and (9)~Metis were
conducted at the Australia Telescope Compact Array (CA)\footnote{The
CA is a part of the Australia Telescope, which is funded by the
Commonwealth of Australia for operation as a National Facility
managed by CSIRO.} on 2004 October 13. The relevant observational
details are summarised in Tables.~\ref{tbl:obslog} and \ref{tbl:obsgeometry}.

\begin{table}[h!tb]
      \caption[]{Technical observing details.}
         \label{tbl:obslog}
\begin{tabular}{lc}
\hline
\hline
      \noalign{\smallskip}
Observatory        & ATCA, Narrabri NSW \\
Longitude          & 149$^{\circ}$32$^{\prime}$56.327$^{\prime\prime}$ E \\
Latitude           &  30$^{\circ}$18$^{\prime}$52.048$^{\prime\prime}$ S \\
Altitude           & 209.3\,m \\
Number/size of antennas          & 5\,$\times$\,22\,m \\
Antenna configuration            & H214C \\
Physical baselines               & 10 (from 82 to 247\,m) \\
UT Date, Time                    & 2004 Oct.\ 13, 07:50-16:30 \\
Precipitable water vapour        & 10-12\,mm \\
Receivers                        & dual SIS \\
System temperatures              & 200-350\,K \\
T$_{sys}$ method                 & chopper wheel \\
                                 & (Kutner \& Ulich \cite{kutner81}) \\
Observing frequency (2 IFs)      & 93.0 and 95.5\,GHz \\
Primary beam FWHM                & 35\,arcsec \\
Pointing uncertainty             & $<$5\,arcsec \\
Total bandwidth                  & 2\,$\times$\,128\,MHz \\
Channels per IF                  & 32 \\
Usable bandwidth                 & 2$\times$ (29\,ch) = 116\,MHz \\
Polarisations                    & 2 orthogonal, linear \\
Sample time                      & 30\,sec \\
Bandpass calibrator              & B1921-293, 10\,min \\
Flux calibrator                  & Uranus, 10\,min \\
Gain (amplitude \& phase) cal.   & B2345-167 \\
Total integration time           & (4)~Vesta:  48\,min \\
                                 & (9)~Metis: 291\,min \\
Projected baselines              & 75\,m to 245\,m \\
                                 & 22 to 75\,k$\lambda$ \\
Synthesised beam                 & $\sim$3\,arcsec \\
Observation cycle & i) point/T$_{sys}$:	  2\,min \\
                  & ii) phasecal B2345-167:   3\,min \\
		  & iii) (4)~Vesta:		  3\,min \\
		  & iv) (9)~Metis:		 19\,min \\
Number/times of cycles                 & 15$\times$ (29\,min 30\,sec) \\
    \noalign{\smallskip}
    \hline
\end{tabular}
\end{table}

\begin{table}[h!tb]
      \caption[]{Observation Geometry at 2004-Oct-13 13:00 UT
                 as seen from the CA site.}
         \label{tbl:obsgeometry}
\begin{tabular}{lllrl}
\hline
\hline
    \noalign{\smallskip}
              & r    & $\Delta$  & $\alpha$	& $\lambda - \lambda_{sun}$ \\
    Asteroid  & [AU] & [AU]	 & [$^{\circ}$] & [$^{\circ}$] \\
      \noalign{\smallskip}
      \hline
      \noalign{\smallskip}
    (4)~Vesta & 2.397416 & 1.525840 & 14.57 & 142.79 \\
    (9)~Metis & 2.276962 & 1.392145 & 14.86 & 144.16 \\
    \noalign{\smallskip}
    \hline
\end{tabular}
\end{table}

At the time of observation, five of the CA's six antennas
had been equipped with receivers for the 3\,mm band (the sixth
antenna was not so equipped because of its fixed location at the
6\,km station, too far to obtain correlated signals at 3\,mm at
this site). 
Because obtaining the most accurate flux densities was our
primary goal, we paid very careful attention to all known
calibration issues. We chose the maximum possible bandwidth
available for each of two intermediate frequencies (IF) in
the local oscillator (LO) chain; the receiver
electronics allowed these two IFs to be up to 2.7\,GHz apart, and so
we chose to centre them at 93.0 and 95.5\,GHz, with a view to
possibly detecting an emissivity variation over this frequency
range (see below). The frequencies in the mid-90s were chosen to
maximise the overall sensitivity of the CA in this
band. The pointing was checked approximately every hour, with
corrections to each antenna's pointing table typically only a few
arcseconds. For bandpass calibration we chose
one of the brightest 3\,mm point sources available in the southern
sky (B1921-293, about 15Jy at this wavelength). We used Uranus
as our absolute flux calibrator since its mm-wave emission appears
to be well-matched by a simple uniform-disc model (Griffin \& Orton
\cite{griffin93}).
This model gives 132.5\,K for Uranus' brightness temperature at
94.25\,GHz, which yields a zero-spacing flux density of 8.6\,Jy
at the epoch of observation. Thus, issues of pointing, bandpass
calibration, or planet-based absolute flux scale calibration
contributed negligibly to the overall error budget (see Table~\ref{tbl:cal}).

\begin{table*}[h!tb]
      \caption[]{Measured Flux Density RMS Errors (per baseline
                 unless otherwise noted).}
         \label{tbl:cal}
\begin{tabular}{lc}
    \hline
    \hline
    \noalign{\smallskip}
RELATIVE UNCERTAINTIES: & \\
Thermal noise in asteroid signal (per cycle, both IFs and polarisations): V, M & 10\%, 15\% \\
Loss due to pointing errors (3\,arcsec) & 0.3\% \\
Bandpass calibration (B1921-293) & 1\% \\
Atmospheric opacity calibration & 2\% \\
Intercycle gain stability (after applying gaincal corrections),  & 2\% \\
Intercycle phase stability (after applying gaincal corrections) & 2$^{\circ}$ (typical) \\
                                                                & 10$^{\circ}$ (extreme)\\
\hline \\
{\bf Net relative uncertainty, per 30\,min cycle, per baseline: V, M} & {\bf 10\%, 15\%} \\
{\bf Net relative uncertainty per cycle, all baselines combined: V, M} & {\bf 3\%, 5\%} (typical) \\
                                                                       & {\bf 20\%} (extreme) \\
\hline \\
ABSOLUTE UNCERTAINTIES: & \\
T$_{sys}$ temperature scale (chopper wheel vs.\ skydip methods), per antenna & 5\% \\
Gain elevation systematic errors & 10\% \\
Uranus model (132.5\,K uniform disc) & 2\% \\
\hline \\
{\bf Total (absolute + relative) uncertainty, all times, all baselines combined} & {\bf 11\%} \\
\multicolumn{1}{r}{per 30\,min cycle} & 25\% (extreme) \\
    \noalign{\smallskip}
    \hline
\end{tabular}
\end{table*}

The main calibration issues at mm wavelengths are the medium-term
($\sim$hours) variability in the complex receiver gains, the atmospheric
opacity, and the antennas' gain vs.\ elevation. Therefore periodic
gain (amplitude and phase) calibration was performed using
B2345-617, which was located 6.5$^{\circ}$ east of (4)~Vesta at the time
of observation. Since (9)~Metis was itself only 2.5$^{\circ}$
north of (4)~Vesta at this time, we were assured of obtaining
an accurate {\it relative} calibration between the quasar and the two 
asteroids, even if the anennas' gain-elevation dependence (see below)
was not well characterised.

From very simple model calculations, we anticipated (4)~Vesta and (9)~Metis would
have 3.2\,mm flux densities on this date of roughly 200 and 20\,mJy,
respectively. In this case we would have ideally achieved similar
signal-to-noise measurements of their rotationally-averaged
brightnesses by integrating on them in time in the inverse-square
of this ratio, or $\sim$1:100. However, a more practical integration-time ratio
turned out to be $\sim$1:6. 
The relative uncertainties of individual cycles are mostly statistical
and we obtained roughly a S/N of 24 for Metis and 42 for Vesta per cycle.
For the global  average of the 15 cycles, the standard errors in the mean (SEM)
do not average down by a factor of 4 because both lightcurves contain a 
number of points at large excursions from the mean, relative to the size of the
error bars. Given the measured flux densities, uncertainties, lightcurve
excursions, an so on, this time ratio serendipitously gave a relative
S/N $\sim$ 35 for both bodies.

To simultaneously
obtain (a) the best calibration, (b) roughly equivalent S/N for
both asteroids, and (c) time-resolved 3.2\,mm lightcurves, we
interleaved observations of our two asteroids in time. We set up
our observe "cycle" as follows: (i) start with a pointing
check and/or T$_{sys}$ calibration, typically taking 2 minutes;
(ii) observe the gain calibrator for 3 minutes; (iii) observe
(4)~Vesta for 3 minutes; and (iv) observe (9)~Metis for 19 minutes.
Each of these cycles thus took close to 30 minutes to execute,
including time for telescope moves.
After starting with the bandpass calibrator and Uranus,
this observe cycle was executed 15 times over the 7.5\,h
span of the observations, finishing with an extra observation
of the gain calibrator.

The post-observation calibration procedure followed standard CA
practices, as implemented in the Miriad software package
(Sault et al.~\cite{sault95}).
The spectral bandpass in both the 93.0 and 95.5\,GHz IFs was
derived to high precision from the B1921-293 data. Variations
in the receiver complex gains and atmospheric opacity and phase
variations were simultaneously removed using B2345-167,
a nearby 3\,mm point source. This step puts B2345-167 and all
the asteroid data on the same relative flux scale, and also
removes any residual effects from inaccuracies in the
gain-elevation correction for each antenna. If this has been
done well, any remaining fluctuations or uncertainties
should be dominated by thermal noise from the
atmosphere + antenna + receiver system, and this indeed seems
to be the case for most of our data (see Table~\ref{tbl:cal},
"relative uncertainties").

However, for a couple of the cycles, the system phase stability
was much more erratic than normal (labelled "extreme" in
Table~\ref{tbl:cal}). In the case of the last cycle (16:00 $<$ UT $<$ 16:30),
this was likely due to the low elevation of our sources
($\sim$30$^{\circ}$), and consequent variability in
atmospheric opacity. For the second cycle (9:45 $<$ UT $<$ 10:15),
its cause was less obvious (being at elevation $\sim$50$^{\circ}$),
but may have been similarly due to short-term atmospheric changes.
Such phase fluctuations have the effect of decorrelating the
visibilities from the source, and reducing their observed amplitudes.
However, the gain calibration seemed to compensate
adequately for this, by boosting the scalar amplitudes of the
visibilities an appropriate amount. The result is that the
vector-averaged visibilities during such episodes are brought
into line with those from more stable cycles. However, we return to this
issue when discussing the asteroids' light curves below.

For absolute calibration, this is first done approximately by
converting the data to the T$_{sys}$ temperature scale. Note that the
data are split up into 4 independent parts, namely the
{\it XX} and {\it YY} correlations for both the 93.0 and 95.5\,GHz IFs. These
are independent of each other since each antenna has two receivers
sensitive to orthogonal linear polarisations from the sources, and
the frontends (which are broadband devices) have their upper and lower IF
signals split in the LO chain and admitted to separate halves of
the correlator. Thus there are four independent T$_{sys}$ measurements
made at the start of each observe cycle. These temperature scale
measurements are thought to be reliable to $\sim$5\%, based on comparisons
between the chopper-wheel method of Kutner \& Ulich (\cite{kutner81}),
which is the standard method used at the CA, and skydip measurements
also performed at the CA by R.\ Sault (private communication).

While the {\it XX} and {\it YY} data for each frequency should be giving us the same
result for an unpolarised source, and this appears to be the case here,
the two IFs may give different amplitudes. This can be due not only
to the receivers' broadband sensitivity as a function of frequency,
but also to the intrinsic spectral index of the source. Thus any differences
between the two IFs are not significant until we have bootstrapped
the data to an absolute flux scale. To disentangle these two effects,
at 3\,mm we observe a source with a known flux density at the
frequencies of interest, such as a planet. Normally, the average of the
corrections from the gain calibrator are applied to the planet data,
which are then compared to a planet model, and a simple, single scaling
factor derived to correct from the relative to the absolute scale.
The correction factor is then applied to the gain calibrator and source data.

However, since the CA's new 3\,mm system had only recently been commissioned
at the time of observation, the correction for each antenna's gain as
a function of elevation above the horizon had not yet been
well-characterised. This is significant since a poor
gain-elevation correction can masquerade as a variation in the
antenna-based receiver gain correction. By itself, this will be
removed by using the gain calibrator to nevertheless give a good
relative flux scale. But, if the flux calibrator is not observed {\it at
the same time and elevation} as the gain calibrator, it will not
be possible to differentiate between a poor gain-elevation correction
and a genuine reciver-gain drift. This can potentially result in
large errors in determining the absolute flux scale, of up to
50\% (I.\ Klamer, private communication), even while the relative
flux scale is quite accurate.

This problem is worst when the flux calibrator is
observed at an elevation which doesn't correspond to any
programme source elevations. In such cases the absolute flux
scale essentially has to be guessed. Fortunately in our case,
Uranus was observed at around 57$^{\circ}$ elevation,
close to the mean elevation of our sources (between 27$^{\circ}$
and 75$^{\circ}$, see also Tbl.~\ref{tbl:tbl3}), which means
correcting for this effect is at
least possible, and moreover should constitute a small effect.
Nevertheless, instead of using the time-averaged gains from
our gain calibrator (i.e.\ data from all elevations) to apply to
Uranus, we used only the corrections from when B2345-167 was very
close to the elevation (57$^{\circ}$) at which we observed Uranus.
This means that whatever the gain-elevation correction {\it should} have
been for our data, any erroneous effect on our absolute flux
scale was completely removed from the data. The only remaining
calibration issue was a roughly 2-hour gap
between the Uranus observations and the commensurate
B2345-167 observations (made before transit), during which
time the receiver gains would have slowly drifted by an unknown
amount. However we also observed B2345-167 about four hours
later (after transit) at the same elevation, where we could
see that the gain corrections had not significantly changed
from the before-transit numbers, to within a $\sim$10\%
level. Based on this, we put the overall gain uncertainty due
to this effect at 10\%, although it may be significantly
less than this.

This uncertainty by far dominates our error
budget for the absolute flux densities (Table~\ref{tbl:cal}).
However, now knowing the nature of the effect, it should be
possible in future projects at the CA to reduce this uncertainty
to perhaps 5\% or less, by a more timely planet observation,
or by several observations spread over a range of elevations.
Despite all this, we emphasise that our relative flux scale
(i.e., between the gain calibrator and the asteroids,
between the two asteroids, or within a single asteroid's data)
is good to $\sim$4\% per cycle (all baselines combined),
or $\sim$1\% overall.

%________________________________________________________________
\section{Observational Results}
\label{sec:obs_res}

\begin{figure}[h!tb]
    \begin{center}
    \rotatebox{270}{\resizebox{!}{\hsize}{\includegraphics{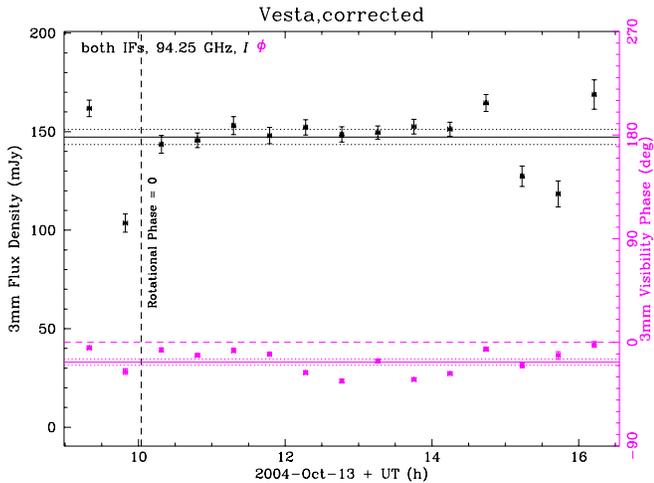}}}
    \caption[]{(4)~Vesta 3.2\,mm flux density $S_I$ of total intensity
     (visibility amplitudes in black, visibility phases in magenta)
     vs.\ UT. Each data point represents a vector
     average of several 30-second samples covering an observing cycle,
     each sample being itself a vector average over (i) 29 $\times$ 4\,MHz
     channels in the correlator, (ii) 2 linear polarisations
     (designated {\it XX} and {\it YY}), (iii) 2 IFs (93.0 and 95.5\,GHz),
     and (iv) 10 baselines, or 1160 correlations per sample. The error
     bars in time simply bracket the range of UT over which the
     samples were taken (5 or 6 samples per cycle in the case of
     (4)~Vesta). The flux error bars are the standard errors in the
     mean (SEMs) from the vector average of the samples, i.e., the rms.\
     of the vector average, divided by the square root of the
     number of correlations that went into the average: the larger
     the errorbars of a plotted point, the more unstable were the data
     that went into that average. The horizontal solid
     and dotted lines represent the mean $\pm$SEM (respectively)
     of all data.}
    \label{fig:fig1}
    \end{center}
\end{figure}

\begin{figure}[h!tb]
    \begin{center}
    \rotatebox{270}{\resizebox{!}{\hsize}{\includegraphics{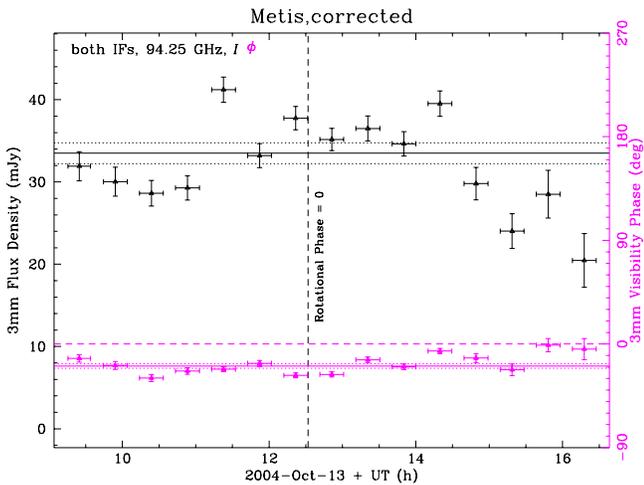}}}
    \caption[]{(9)~Metis 3.2\,mm $S_I$ vs.\ UT, otherwise the same
     as Fig.~\ref{fig:fig1}. The (9)~Metis data have about 38 samples per cycle, 
     hence the larger time error bars than in Fig.~\ref{fig:fig1}.
     }
    \label{fig:fig2}
    \end{center}
\end{figure}

We present our flux density measurements as a function of
UT for (4)~Vesta and (9)~Metis in Figs.~\ref{fig:fig1} and \ref{fig:fig2}
and Table~\ref{tbl:tbl3}.

\begin{table}[h!tb]
  \begin{center}
  \caption[]{ATCA calibrated flux densities for an average frequency
   of 94.25\,GHz (3.18\,mm) as a function of time. The corresponding
   rotational phases are also given (based on rotation periods of
   5.34\,h for (4)~Vesta and 5.08\,h for (9)~Metis together with specific
   zero points, see Sect.~\ref{sec:tpm}). Note, that the errors given
   are the relative uncertainties which are dominated by the
   thermal noise. For a better judgement of the errorbars, the observed
   elevations are also given.}
  \label{tbl:tbl3}
     \begin{tabular}{cccrr}
     \hline \hline 
     \noalign{\smallskip}
                  &  FD   & Error  & Rotational         & Obs.\ Elev.\ \\
     Julian Date  &  [mJy] & [mJy] & phase [$^{\circ}$] & [$^{\circ}$] \\
     \noalign{\smallskip}
     \hline 
     \noalign{\smallskip}
     \multicolumn{5}{l}{(4)~Vesta} \\
     \noalign{\smallskip}
     \hline 
     \noalign{\smallskip}
    2453291.88872 & 161.8 &  4.2 & 312.38 & 51.4 \\
    2453291.90920 & 103.6 &  4.6 & 345.51 & 57.4 \\
    2453291.92969 & 143.6 &  4.5 &  18.64 & 63.4 \\
    2453291.95017 & 145.6 &  3.8 &  51.77 & 68.4 \\
    2453291.97066 & 153.1 &  4.6 &  84.91 & 72.5 \\
    2453291.99115 & 148.0 &  4.2 & 118.05 & 74.6 \\
    2453292.01163 & 152.2 &  3.9 & 151.17 & 74.0 \\
    2453292.03212 & 148.6 &  3.9 & 184.31 & 70.8 \\
    2453292.05260 & 149.6 &  3.4 & 217.43 & 66.1 \\
    2453292.07309 & 152.6 &  3.8 & 250.57 & 60.6 \\
    2453292.09358 & 151.3 &  3.6 & 283.71 & 54.7 \\
    2453292.11406 & 164.6 &  4.3 & 316.83 & 48.6 \\
    2453292.13455 & 127.4 &  5.2 & 349.97 & 42.4 \\
    2453292.15503 & 118.4 &  6.6 &  23.09 & 36.1 \\
    2453292.17552 & 168.9 &  7.5 &  56.23 & 29.7 \\
     \noalign{\smallskip}
     \hline 
     \noalign{\smallskip}
     \multicolumn{5}{l}{(9)~Metis} \\
     \noalign{\smallskip}
     \hline 
     \noalign{\smallskip}
    2453291.89219 &  31.9 &  1.8 & 138.99 & 51.2 \\
    2453291.91267 &  30.0 &  1.8 & 173.82 & 57.1 \\
    2453291.93316 &  28.6 &  1.5 & 208.68 & 62.6 \\
    2453291.95365 &  29.3 &  1.5 & 243.53 & 67.3 \\
    2453291.97413 &  41.2 &  1.5 & 278.37 & 70.8 \\
    2453291.99462 &  33.2 &  1.5 & 313.23 & 72.3 \\
    2453292.01510 &  37.8 &  1.4 & 348.06 & 71.2 \\
    2453292.03559 &  35.2 &  1.4 &  22.92 & 68.0 \\
    2453292.05608 &  36.5 &  1.5 &  57.77 & 63.4 \\
    2453292.07656 &  34.6 &  1.5 &  92.61 & 58.0 \\
    2453292.09705 &  39.5 &  1.6 & 127.47 & 52.2 \\
    2453292.11753 &  29.8 &  2.0 & 162.30 & 46.2 \\
    2453292.13802 &  24.0 &  2.1 & 197.16 & 40.0 \\
    2453292.15851 &  28.5 &  2.9 & 232.01 & 33.7 \\
    2453292.17899 &  20.5 &  3.3 & 266.85 & 27.4 \\
     \noalign{\smallskip}
     \hline 
     \end{tabular}
  \end{center}
\end{table}

\begin{table}[h!tb]
  \begin{center}
  \caption[]{ATCA calibrated flux densities, spectral index, colour
   (rotationally averaged). The errors given correspond the statistical
   uncertainties. As specified in Table~\ref{tbl:cal}, the total absolute flux
   error is around 11\%. The coefficients are specified for the sinusoidal fit 
   $S(\phi) = S_0 [ 1 + A sin 2 \pi (\phi - \phi_0)]$.
}
  \label{tbl:tbl4}
     \begin{tabular}{ccc}
     \hline \hline 
     \noalign{\smallskip}
   & (4)~Vesta & (9)~Metis \\
   \noalign{\smallskip} \hline \noalign{\smallskip}
   $S_0 \pm SEM$ \hspace{1cm} all data & 147.3 $\pm$ 3.8 mJy & 33.5 $\pm$ 1.0 mJy \\
          \multicolumn{1}{r}{93.0 GHz} & 147.3 $\pm$ 3.8 mJy & 32.4 $\pm$ 1.1 mJy \\
          \multicolumn{1}{r}{95.5 GHz} & 147.5 $\pm$ 4.8 mJy & 34.6 $\pm$ 1.2 mJy \\
           Sinusoidal fit: $A \pm SEM$ & $<$ $\pm$ 5\%       & 12\% $\pm$ 3\% \\
  \multicolumn{1}{r}{$\phi_0 \pm SEM$} & ---                 & -0.175$\pm$0.05 \\
     Spectral index $\alpha$ $\pm$ SEM & 0.1 $\pm$ 1.6       & 2.5 $\pm$ 1.8 \\
	     3.2\,mm "colour" $\pm$ SEM & -1.9 $\pm$ 1.6      & +0.5 $\pm$ 1.8 \\
              M:V flux ratio $\pm$ SEM & \multicolumn{2}{c}{0.224 $\pm$ 0.009} \\
     \noalign{\smallskip}
     \hline 
     \end{tabular}
  \end{center}
\end{table}

For large main-belt asteroids the Standard Thermal Model (STM,
Lebofsky et al.\ \cite{lebofsky86}) is generally used
to estimate flux densities. Applying the STM with the default
parameters for beaming, emissivity and phase angle corrections
resulted in a flux of 212\,mJy for (4)~Vesta and 22.6\,mJy
for (9)~Metis using the best published size and albedo values
(see Sect.~\ref{sec:tpm} for the detailed values and the references).
The first remarkable result of our data is that the 3.2\,mm flux density of
(4)~Vesta is about 30\% less, and of (9)~Metis roughly 50\% higher, than
that expected from the simple STM predictions.
Given the discussion above about relative and absolute flux calibration,
we regard these results as fairly robust. For example, if the absolute
flux scale is systematically wrong (due, e.g., to a gain drift between
the Uranus and B2345-167 observations which was larger than the
10\% assumed), it is conceivable that one or other of the asteroids'
flux densities could be brought into line with the prediction from
the STM, but not both at once. Thus the mean $\pm$ SEM (standard errors in the mean)
3.2\,mm flux ratio (9)~Metis/(4)~Vesta is more than double the prediction of
the STM of a ratio $\sim$0.1 (see Table~\ref{tbl:tbl4}).

\begin{figure}[h!tb]
    \begin{center}
    \rotatebox{270}{\resizebox{!}{\hsize}{\includegraphics{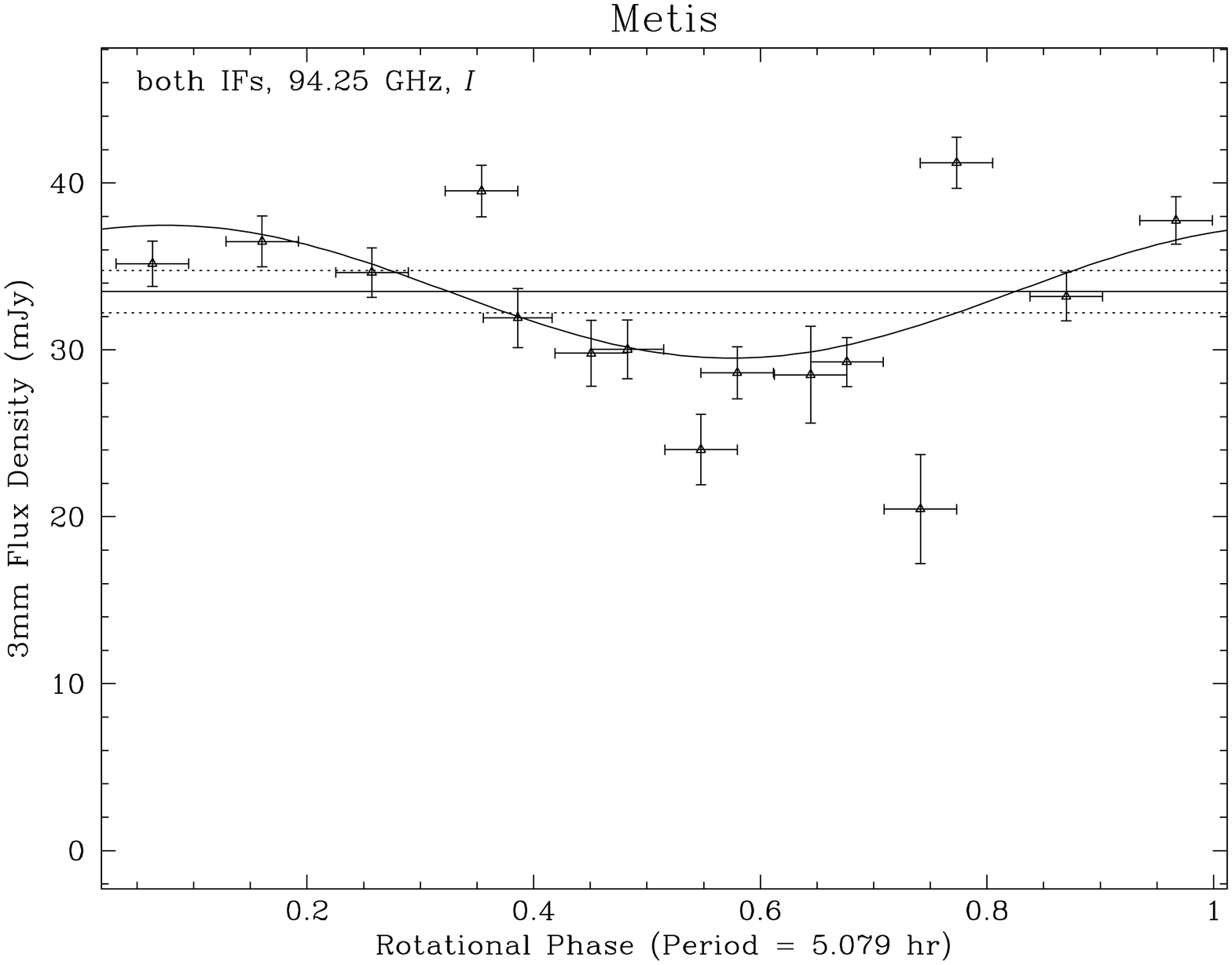}}}
    \caption[]{(9)~Metis total intensity I vs rotational phase $\phi$.
               The curve is for the best fit sinusoid
               described in the text. In Figs.~\ref{fig:fig3} and \ref{fig:fig4}, the
	       zero of $\phi$ is taken at 2004-Oct-13 12:31:51 UT.
	       The other features are the same as in Fig.~\ref{fig:fig1}}
    \label{fig:fig3}
    \end{center}
\end{figure}

\begin{figure}[h!tb]
    \begin{center}
    \rotatebox{270}{\resizebox{!}{\hsize}{\includegraphics{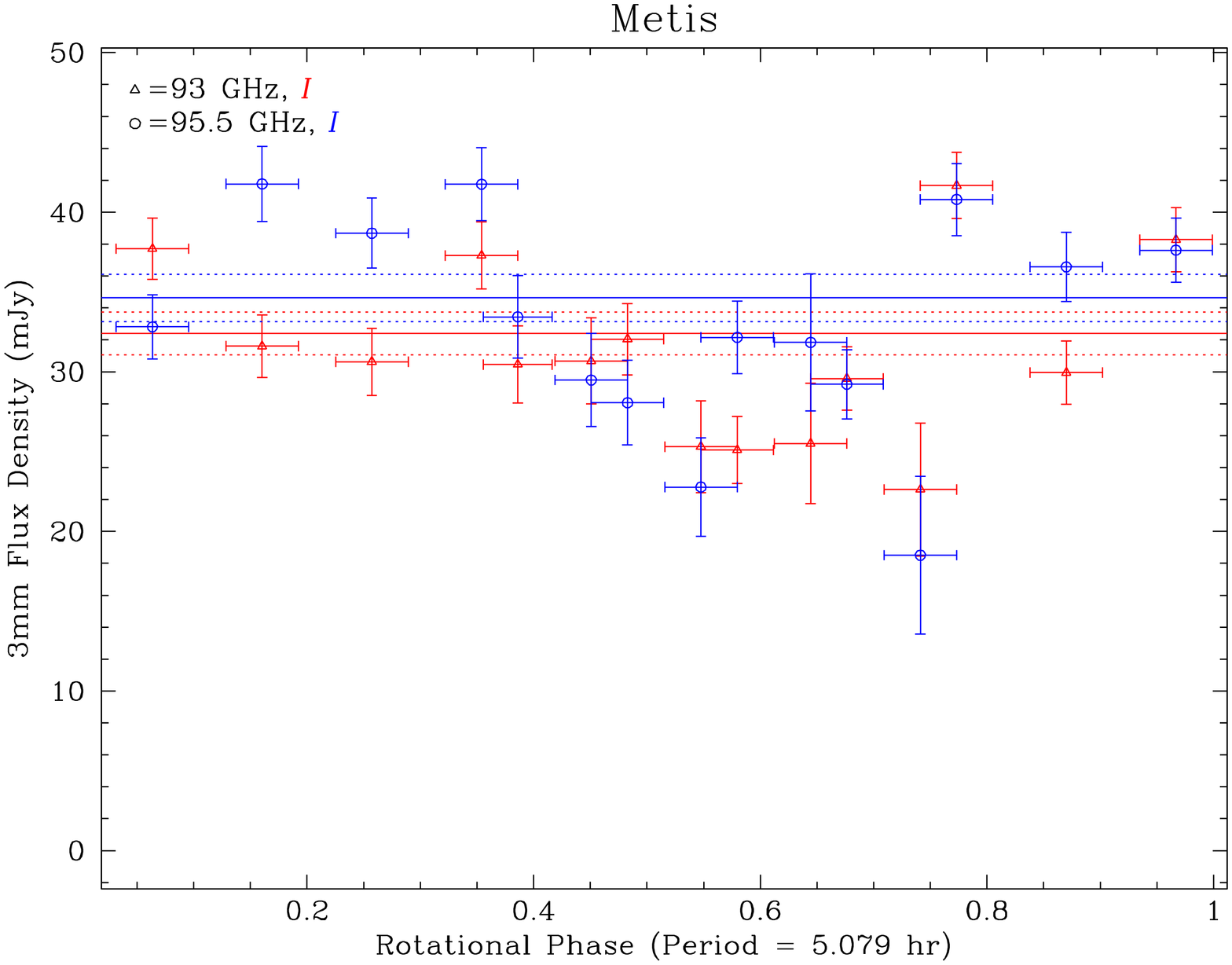}}}
    \rotatebox{270}{\resizebox{!}{\hsize}{\includegraphics{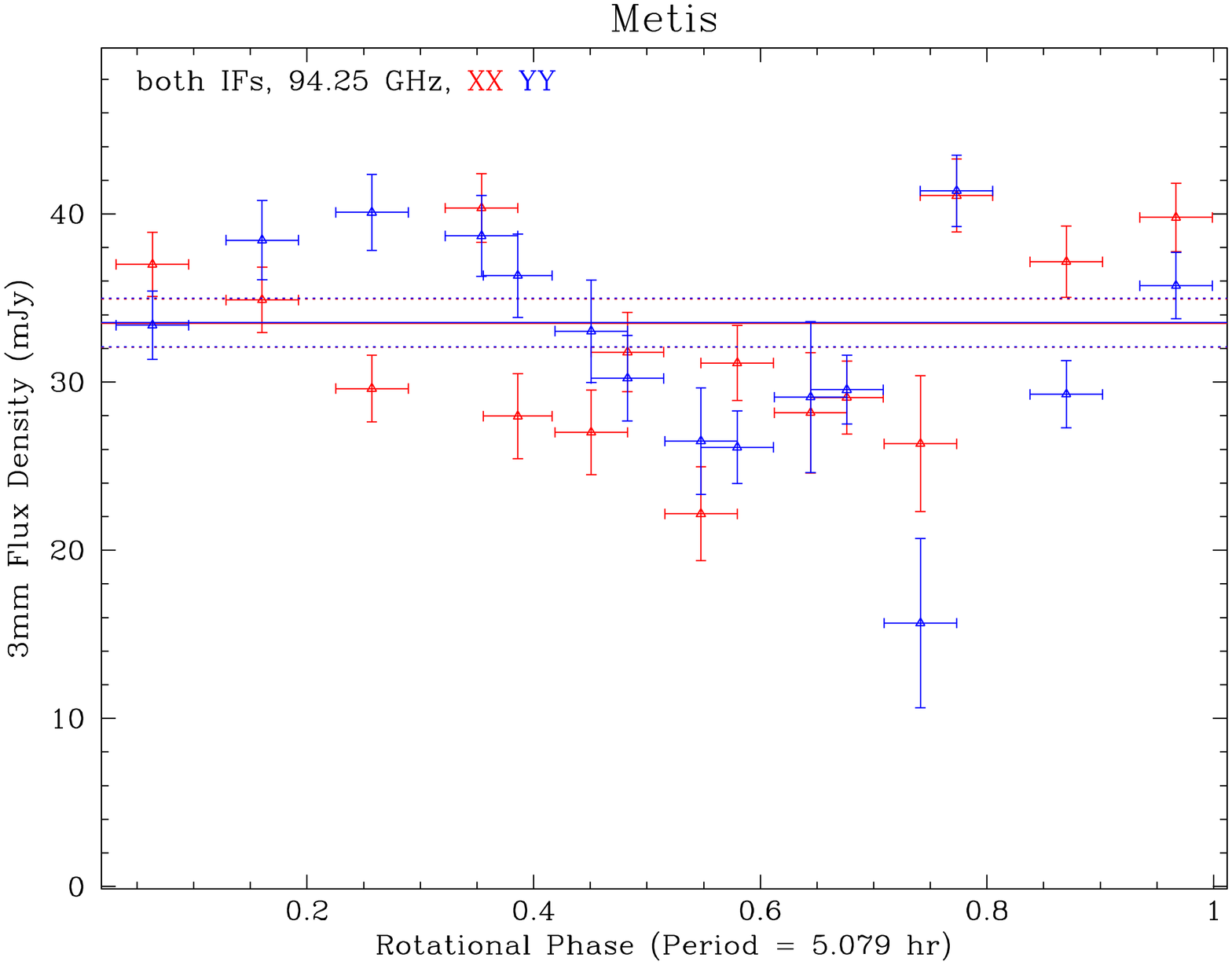}}}
    \rotatebox{270}{\resizebox{!}{\hsize}{\includegraphics{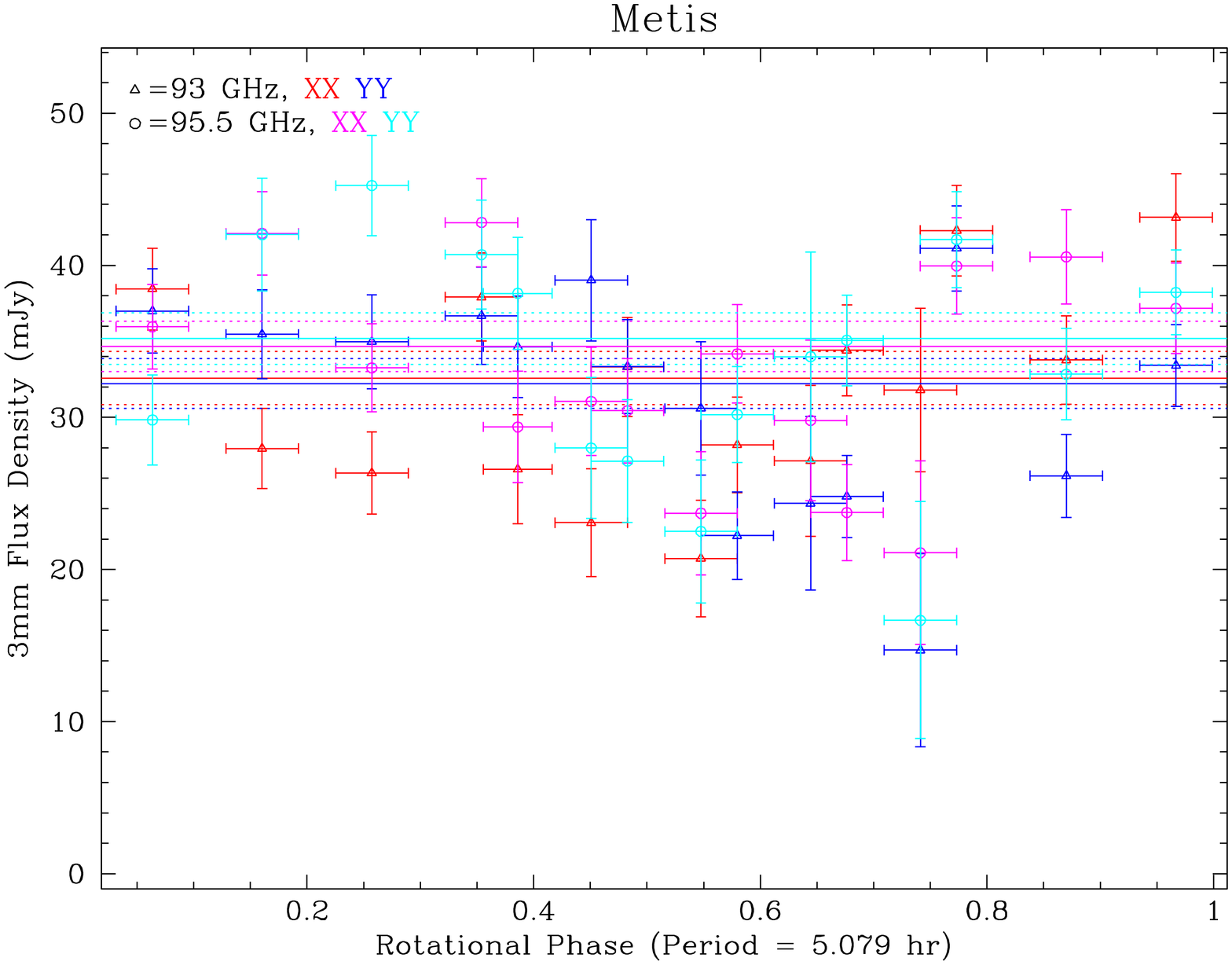}}}
    \caption[]{Subsets of the data from Fig.~\ref{fig:fig3}. (a) (9)~Metis I-93
               and I-95.5 GHz vs $\phi$; each frequency is
	       averaged over both polarisations. (b) (9)~Metis {\it XX} and {\it YY}
	       vs $\phi$; each polarisation is averaged over both
	       frequencies. (c) (9)~Metis data vs $\phi$; each
	       frequency-polarisation subset is split out separately.}
    \label{fig:fig4}
    \end{center}
\end{figure}

\begin{figure}[h!tb]
    \begin{center}
    \rotatebox{270}{\resizebox{!}{\hsize}{\includegraphics{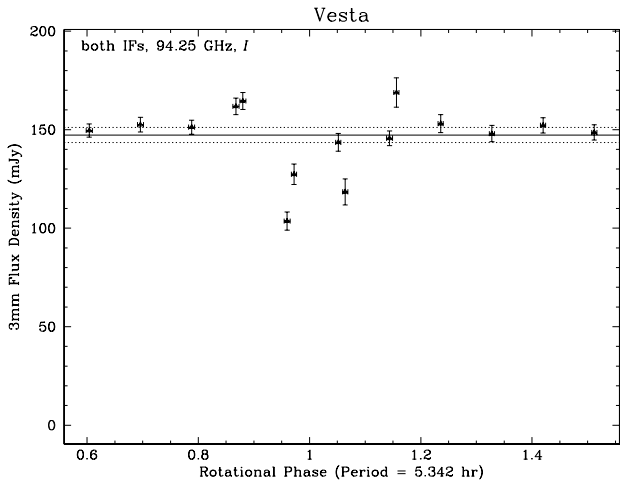}}}
    \caption[]{(4)~Vesta I vs $\phi$, otherwise the same as for Fig.~\ref{fig:fig3}.
               For (4)~Vesta the zero of $\phi$ is taken at 
	       2004-Oct-13 10:02:10 UT (here and also in Fig.~\ref{fig:fig6}).}
    \label{fig:fig5}
    \end{center}
\end{figure}

\begin{figure}[h!tb]
    \begin{center}
    \rotatebox{270}{\resizebox{!}{\hsize}{\includegraphics{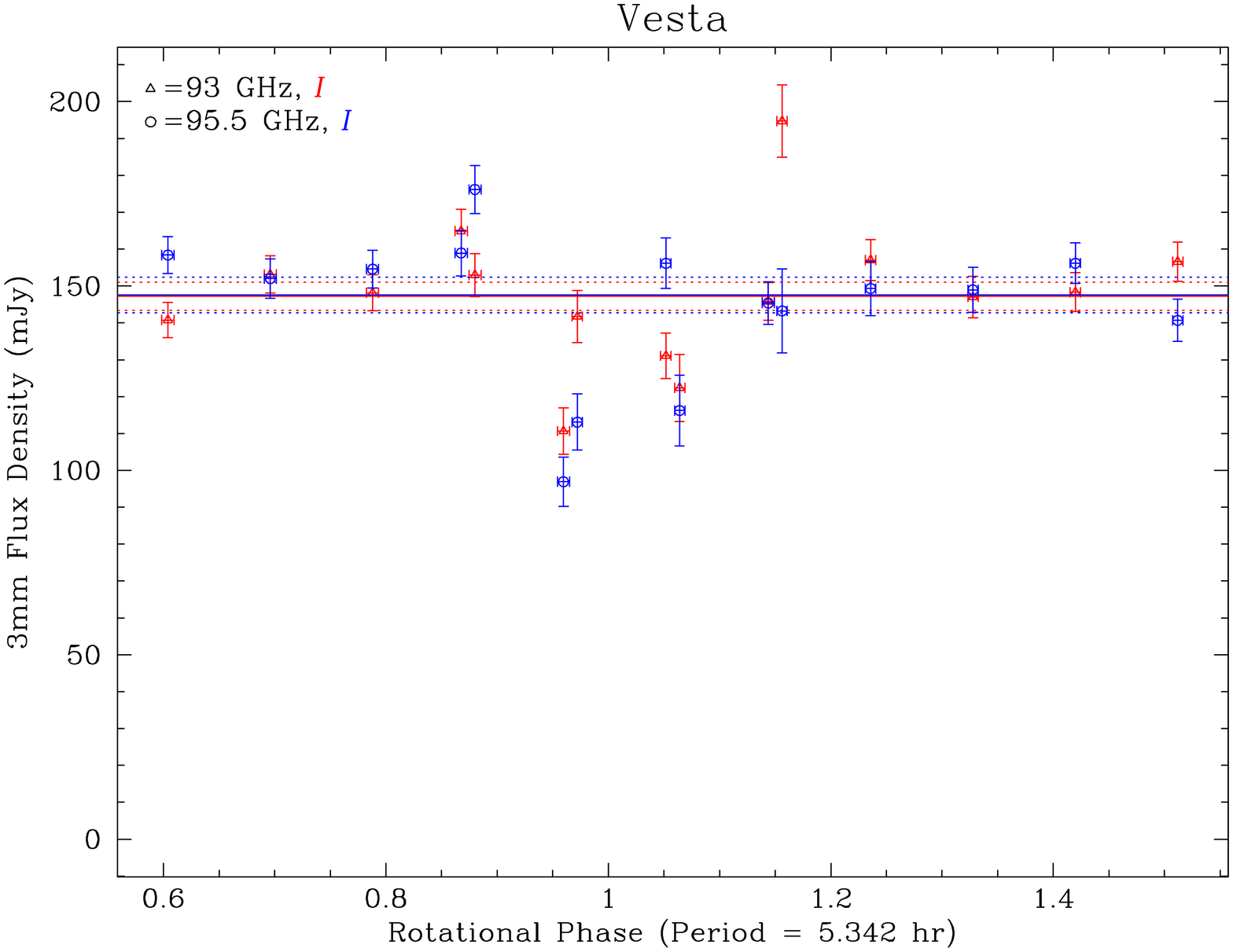}}}
    \rotatebox{270}{\resizebox{!}{\hsize}{\includegraphics{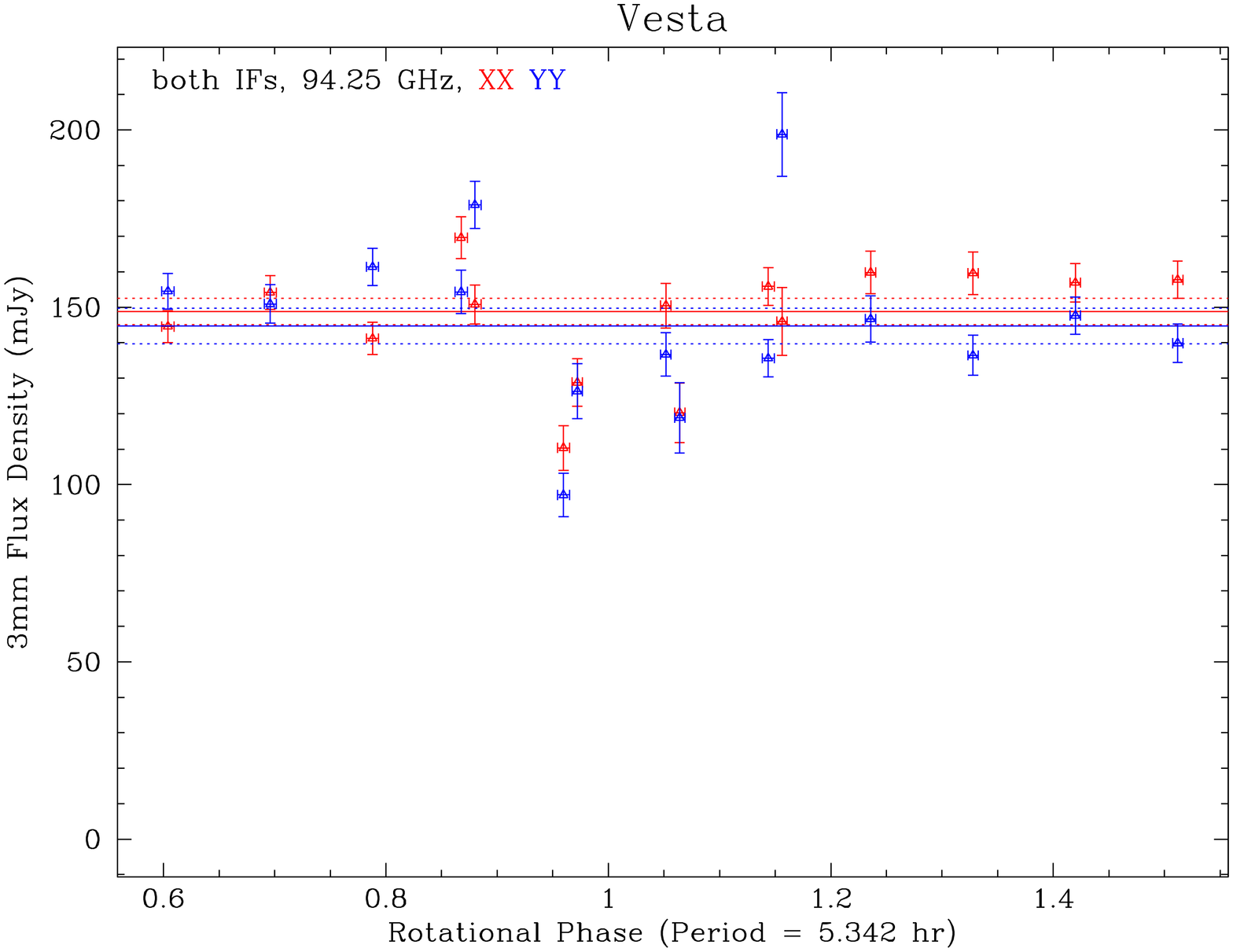}}}
    \rotatebox{270}{\resizebox{!}{\hsize}{\includegraphics{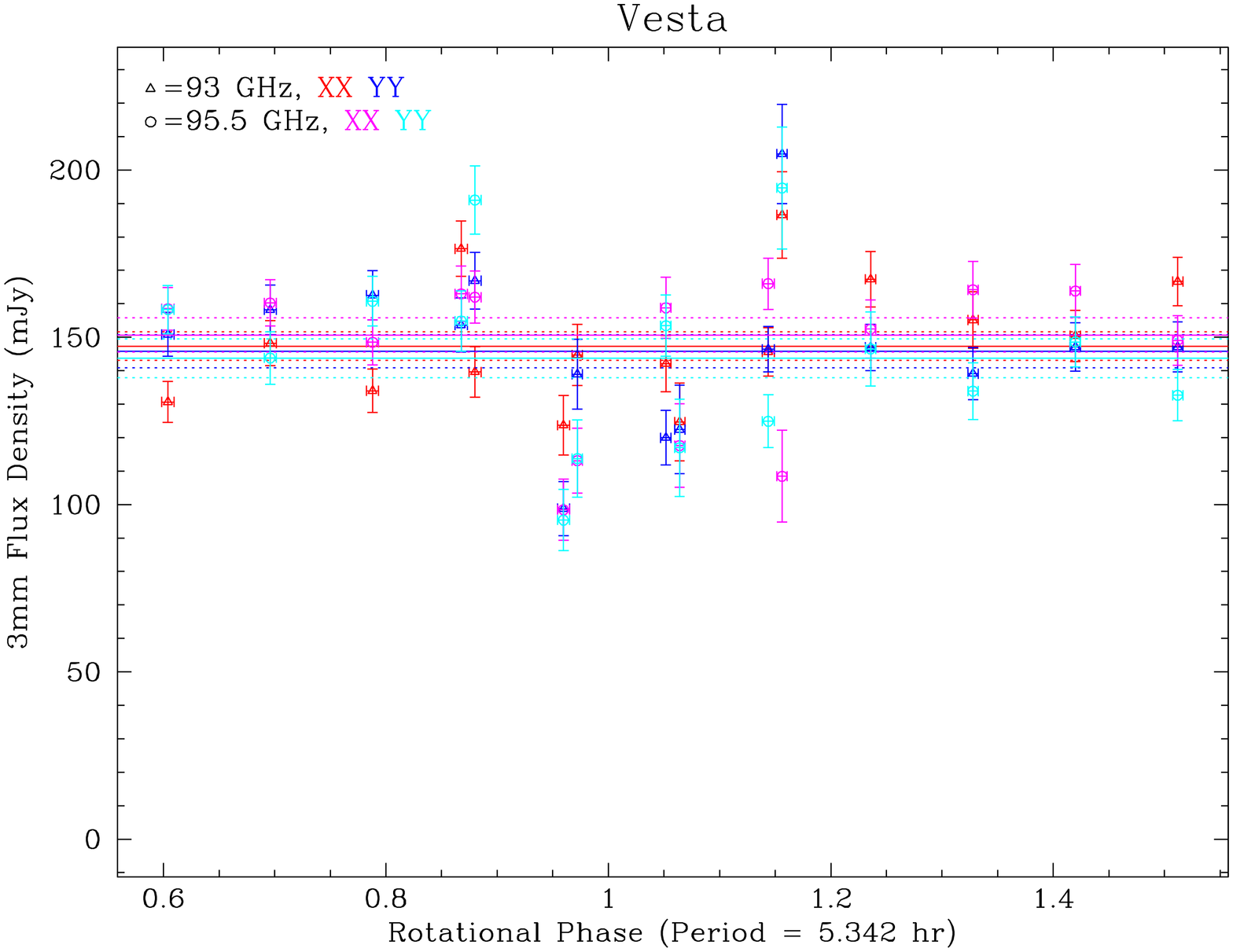}}}
    \caption[]{Subsets of the (4)~Vesta data from Fig.~\ref{fig:fig5} vs $\phi$,
               otherwise the same as for Fig.~\ref{fig:fig4}.}
    \label{fig:fig6}
    \end{center}
\end{figure}

Upon further inspection, there is a suggestion in
the data of significant 3.2\,mm lightcurves for both asteroids.
In (9)~Metis' case, the variation over the asteroid's sidereal period
of 5.079\,h appears to be large ($\pm$12\% around the mean) and
sinusoidal (see Fig.~\ref{fig:fig3}). This suggests a significant
hemispherical difference in the 3.2\,mm emissivity of this body.
The sinusoid seems to persist in whatever division of the data
we plot: total intensity (Fig.~\ref{fig:fig3}), or split by
polarisation, frequency, or both (Fig.~\ref{fig:fig4} a-c).
The sinusoid also seems to be reproduced by data
taken at different UT but the same rotational phase. In light of
the small relative and absolute uncertainties in the flux
density scale, all these features strongly suggest that the
effect is real. The parameters of the sinusoidal fit, of the form

\begin{equation}
S(\phi) = S_0 [ 1 + A sin 2 \pi (\phi - \phi_0)]
\end{equation}

\noindent
(where S is the flux density at rotational phase $\phi$
[measured in units of the rotational period P], $S_0$ is
the rotationally-averaged mean flux density, A is the
amplitude of the sinusoid, and $\phi_0$ is the phase offset
[also measured in units of P]), are given in Table~\ref{tbl:tbl4}.
All results quoted in Table~\ref{tbl:tbl4} are from weighted
least-squares fits to the individual cycle data points, and
where the weight for each point was taken as its $(rms)^{-2}$.

%\subsection{(4)~Vesta}

For (4)~Vesta, we also see lightcurve variations over this body's 5.342\,h
sidereal period. A strong dip $\sim$25\% below the mean flux density
(near $\phi$ $\sim$0.95) appears in total intensity
(Fig.~\ref{fig:fig5}). The dip becomes somewhat less
convincing in the frequency- or polarisation-split data
(Fig.~\ref{fig:fig6} a-c), but is still present. Bright spikes
in (4)~Vesta's lightcurve to either side
of the dip (near $\phi$ $\sim$0.87 and 1.15) can also be seen. These features
do seem to repeat from one rotation to the next, and are not
confined to only those cycles which experienced extreme
phase instability\footnote{The larger the errorbars of a plotted
point, the more unstable were the data that went into that
average.} (see Section~\ref{sec:obs_cal}).
This suggests the dip and spikes may not simply be artefacts
of anomalous phase errors in some of the data. But since their
existence is not easy to explain, their reality should be confirmed
by future experiments. Parenthetically, it is worth noting
that for the flat part of (4)~Vesta's lightcurve (10:15 $<$ UT $<$ 14:15,
Fig.~\ref{fig:fig1}; or 0.2 $<$ $\phi$ $<$ 0.8, Fig.~\ref{fig:fig5}),
the fluctuations are entirely consistent with thermal noise.

%\subsection{(9)~Metis}

Returning to (9)~Metis, we see some similarly interesting deviations
from the fit sine curve. Most notably, at $\phi$ $\sim$0.75 data
from two different rotations appear completely inconsistent.
The lower point at $\sim$20\,mJy might be easily explained away
as an artefact of the extreme phase variations during the last
cycle (UT $\sim$16:20), insufficiently calibrated.
However the upper point at $\sim$41\,mJy (UT $\sim$11:20, cycle 5)
is not so easily dismissed. The separate data for this point in
Fig.~\ref{fig:fig6}c are very tightly grouped, giving a point
in total intensity (Fig.~\ref{fig:fig5}) standing $\sim$ 5$\sigma$
above the fitted sine curve. So while this makes the spike
appear real, it is hard to imagine a surface feature that could give
rise to it. Perhaps more likely is a $\sim$10\,min long atmospheric
phase decorrelation episode during the gaincal measurement before cycle
5, and during the cycle-5 measurement for (4)~Vesta, which then ended
for the cycle-5 measurement of (9)~Metis. This would result in a
correctly-compensated cycle-5 value for (4)~Vesta, but an over-corrected
cycle-5 value for (9)~Metis. Whether or not this explanation is correct,
averaging in the low cycle-15 point to the high cycle-5 point for 
(9)~Metis would bring the mean of the two points at $\phi \sim$ 0.75 
very close to the fitted sine.
%Averaging in the low cycle-15 point, however, would
%bring the mean of the two very close to the fitted sine.
Either way, without these two cycles, the rest of the (9)~Metis data
are again consistent with thermal noise around the sine curve.

%\subsection{General aspects}

A third interesting result for these bodies is the evidence
in the data for different spectral indices at 3.2\,mm (or equivalently,
emissivity variations, or "3.2\,mm-colours").
Despite the issues with the absolute flux scale discussed above, the
relative calibration between the two IFs at 93.0 and 95.5\,GHz
should be good to the level in Table~\ref{tbl:cal} quoted for
the relative flux scale, since whatever effect receiver-gain
drifts and the gain-elevation correction have on our data,
they are likely to be very similar for the two IFs. We define the
spectral index $\alpha$ between two frequencies by
$ S(\nu) \sim \nu^{\alpha} $;
then a roughly 200\,K blackbody (or grey-body) should have
$\alpha = 2.0$ at these frequencies, being in the Rayleigh-Jeans
tail of the Planck function. This allows us to also define
the "3.2\,mm colour" as $\alpha -2$. From Figs.~\ref{fig:fig4}a
and \ref{fig:fig6}a we obtain the values for $\alpha$ and the
3.2\,mm colour in Table~\ref{tbl:tbl4}. Put another way, (9)~Metis'
3.2\,mm colour is consistent with a grey body, while (4)~Vesta's
3.2\,mm colour is much "redder" than a grey body. The
uncertainties are fairly large, however, due to the small frequency
range (3\%), so these results must be treated as fairly tentative,
although the appearance of different colours for the two asteroids
is unlikely to be due to a {\it systematic} instrumental effect.
Moreover, the errors for the spectral index quoted in
Table~\ref{tbl:tbl4} overstate the uncertainty. Omitting the
Olbers-related data points (see below) from Vesta's lightcurve,
these SEMs are reduced by a factor $\sim$3. If confirmed, these
3.2\,mm colours suggest that while (9)~Metis' emissivity is
roughly constant around 3\,mm, (4)~Vesta's emissivity might actually
be rising towards longer wavelengths. We return to this issue
in Sect.~\ref{sec:discussion}.

In contrast, we note there {\it is} a residual instrumental effect
visible in the polarisation split data (Figs.~\ref{fig:fig4}b, \ref{fig:fig6}b).
For both (4)~Vesta and (9)~Metis at both frequencies, the {\it XX} data appear to be
systematically brighter than the {\it YY} data before transit, and systematically
fainter after transit. The difference in all cases is at a level
of $\pm$5\%, and seems to be related to a similar effect appearing also
in the gain calibrator data. For B2345-167, the {\it XX}-{\it YY} difference before
and after transit was around $\pm$15\%, although this was calibrated
out as part of the normal data reduction. This similarity might be
explained, for example, if there was a real difference in the
gain-elevation correction for the two polarisations, perhaps as a
result of some differential flexure in the optical path for the
two receivers as the antennas tracked the calibrator and
sources at slightly different elevations. This is a fairly unlikely
explanation, however, and is more likely just due to the fact that
the instrumental polarisation at 3\,mm had not yet been determined
at the time of our observations. With proper polarisation calibration,
future experiments at the CA may well enable all the Stokes
parameters I, Q, U, and V to be measured at 3\,mm.

Finally, we point out that the asteroid phases (Figs.~\ref{fig:fig1} and
\ref{fig:fig2}) centre around -17$^{\circ}$ and
-20$^{\circ}$ respectively for (4)~Vesta and (9)~Metis.
%This is likely not due to any instrumental or miscalibration effects, but is rather a
%testament to how well the CA's 3\,mm system worked.
It is most probable that these non-zero phases indicate a shift in each
asteroid's position from the ephemeris-determined phase-pointing centre.
With synthesised beams around 3$^{\prime \prime}$, the positional
shifts corresponding to these phases would be around a tenth of this figure,
or 0.3$^{\prime \prime}$.
This angle also roughly corresponds to each asteroid's
subtended diameter as viewed from Earth at the time of observation
(0.238$^{\prime \prime}$ for (4)~Vesta, 0.170$^{\prime \prime}$ for (9)~Metis).
Ephemeris uncertainties are even smaller, about a tenth of the diameters.

In the absence of more reasonable explanations, these non-zero visibility
phases are possibly due to a combination of two effects:
a hint of a 24-hr variation about the zero visibility phase due to a slight
error in the baseline solution, and in Vesta's case at least, a 5-hr variation
about the mean visibility phase. This number, if real, seems inescapably
connected to the body itself. It is as if bright or dark features are
rotating into and out of view about the centre of the body, giving an
apparent shift in the mean position of (4)~Vesta, lending further credence
to the interpretation of the lightcurve variations as being due to surface
emissivity features (see Sect.~\ref{sec:discussion}).
If improvements to the CA's performance at 3mm, such as implementing
some kind of phase-tracking scheme, can be
made in the future, the possibility exists of directly imaging at
3\,mm some of the larger asteroids at an apparition's perigee
using baselines $\sim$3\,km. For asteroids such as Ceres or
Pallas, this would allow spatially resolved modelling of the
regolith properties over their surfaces.

%\clearpage
%\newpage
%________________________________________________________________
\section{Thermophysical modelling}
\label{sec:tpm}

A recent thermophysical model (TPM) describes the thermal emission
of asteroids, including size, shape, albedo, rotational, surface
regolith and thermal behaviour aspects (Lagerros \cite{lagerros96};
\cite{lagerros97}; \cite{lagerros98}).
We used this TPM together with the best available shape models
and spin vector solutions (see tables~\ref{tbl:tpm_vesta}
and \ref{tbl:tpm_metis}).
The values of the thermal properties are taken from
M\&L 1998, 2002, with a wavelength-dependent
emissivity model, a thermal inertia of
15\,J\,m$^{-2}$\,s$^{-0.5}$\,K$^{-1}$ and a "default
beaming model" with $\rho=0.7$ (the r.m.s.\ of the surface slopes)
and $f=0.6$ (the fraction of the surface covered by craters).
The TPM beaming model accounts for the non-isotropic heat radiation,
noticeable at phase angles close to opposition.
The most critical parameters for absolute flux predictions
at thermal wavelengths are the effective diameter and the albedo
values of an asteroid and its emissivity (e.g., M\"uller \cite{mueller02a}).
They are discussed in more details in the subsequent sections.

\subsection{Asteroid (4)~Vesta}

\begin{table}[h!tb]
  \begin{center}
  \caption[]{Thermophysical model input parameter and resulting flux
             densities. Note: The implementation of the Thomas
             et al.\ (\cite{thomas97}) shape model produces a zero
             rotational phase at 2004-Oct-13 10:02:10 UT for
             the given spin vector (SV) and rotation period. }
    \label{tbl:tpm_vesta}
  \begin{tabular}{ll}
  \hline
  \hline
  \noalign{\smallskip}
   \multicolumn{2}{c}{(4)~Vesta} \\
   \noalign{\smallskip} \hline \noalign{\smallskip}
  H, G:  &  3.20\,mag, 0.34 (M\&L 1998, refs.\ therein) \\
  shape: & Thomas et al. 1997 (HST observations) \\
  SV and zeropoints:   &  Thomas et al. 1997 (note added in proof) \\
  Rot.\ period:        & 0.2225887\,days (Drummond et al.\ \cite{drummond88}) \\
  $\epsilon$-model:    & $\epsilon=f(\lambda)$ (M\&L 1998) \\
  thermal inertia $\Gamma$ & 15\,J\,m$^{-2}$\,s$^{-0.5}$\,K$^{-1}$  (M\"uller et al.\ \cite{mueller99})\\
  beaming parameter & $f=0.6$, $\rho=0.7$ (M\&L 2002) \\
   \noalign{\smallskip} \hline \hline \noalign{\smallskip}
   \multicolumn{2}{c}{TPM Predictions} \\
   \noalign{\smallskip} \hline \noalign{\smallskip}
  FD at 3180\,$\mu$m   & 170$\pm$8\,mJy ($\epsilon (\lambda)$, M\&L 1998)\\
  at observation epoch & 146$\pm$8\,mJy ($\epsilon (\lambda)$, but with \\
  (rotation averaged)  & $\epsilon \sim$ 0.6 at submm/mm)\\
   \noalign{\smallskip} \hline \noalign{\smallskip}
  mm-lc amplitude & $\pm$ 4\,mJy (= $\pm$ 2-3\%) \\
  TPM prediction  & (shape-dominated) \\
   \noalign{\smallskip} \hline \noalign{\smallskip}
  \end{tabular}
  \end{center}
\end{table}

\begin{figure}[h!tb]
    \begin{center}
    \resizebox{\hsize}{!}{\includegraphics*{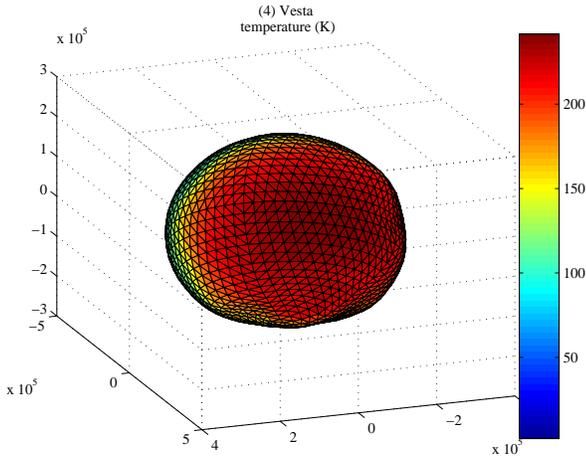}}
    \caption[]{Temperature distribution on the surface
               of (4)~Vesta, based on the HST-shape model in combination with
              the thermophysical parameters and calculated
	      for 2004-Oct-13 09:00:00 UT.}
    \label{fig:vesta_tpm}
    \end{center}
\end{figure}

Thomas et al.\ (\cite{thomas97}) derived the size of (4)~Vesta from
a series of HST images. The best fit was obtained by a triaxial
ellipsoid of radii 289, 280 and 229\,km (all $\pm$ 5\,km),
specified in "note added in proof". This corresponds to an
effective diameter of 529.2\,km. Our best thermophysical model
solution, based on more than 200 independent thermal observations
in the mid-IR to mm-range, resulted in a slightly larger
value of 541.0\,km. The standard deviation of the radiometric
solution is 28.8\,km which brings both size values into
agreement. Tedesco (\cite{tedesco89}) derived an average geometric albedo of 0.38
at 0.55\,$\mu$m. From the most recent albedo value (p$_\mathrm{H}$=0.370)
by Shevchenko \& Tedesco (\cite{shevchenko06}), we calculated a radiometric
value of p$_\mathrm{V}$ = 0.33$\pm$0.04, again in agreement within the 
errorbars. Binzel et al. (\cite{binzel97}) reported a low albedo
feature on the surface (which they proposed to name "Olbers")
with an estimated value of 0.3 at 0.439\,$\mu$m, 
about 10-20\% lower than the average at this wavelength. 

We take the above full range of diameter/albedo
values into consideration for the further analysis.
We applied a (4)~Vesta-specific wavelength-dependent emissivity as determined
by M\&L (\cite{mueller98}) and in agreement with 
Redman et al. (\cite{redman92}). This emissivity model has values of around
0.6 in the submm-range and slightly larger values of around 0.7 at
millimetre-range (see also Webster \& Johnston \cite{webster89} for the
slightly higher values at cm-wavelength).
The translation into a temperature picture is shown in Fig.~\ref{fig:vesta_tpm} for the
specific observing and illumination geometry during the CA observations.

\subsection{Asteroid (9)~Metis}

\begin{table}[h!tb]
  \begin{center}
  \caption[]{Thermophysical model input parameter and resulting flux densities.Note:
             Our implementation of the Kaasalainen (priv.\ comm.) shape model
	     produces a zero rotational phase at 2004-Oct-13 12:31:51 UT.}
    \label{tbl:tpm_metis}
  \begin{tabular}{ll}
  \hline
  \hline
  \noalign{\smallskip}
   \multicolumn{2}{c}{(9)~Metis} \\
   \noalign{\smallskip} \hline \noalign{\smallskip}
  H, G  & 6.28\,mag, 0.17  (Lagerkvist et al.\ \cite{lagerkvist01}) \\
  shape & 2040 trishape surface elements \\
        & and 1022 vertices (Torppa et al.\ 2003) \\
  Spin vector      & $\beta_P$, $\lambda_P$, P$_{sid}$ (hrs): \\
  (Marchis et al.\ 2006) &  21.17, 180.48, 5.07917628 \\
  Zeropoints            &  T$_0$,   $\phi_0$ \\
  (MK, priv.\ comm.)        &  2433222.66230, 270.0 \\
  $\epsilon$-model:    & $\epsilon=f(\lambda)$ (M\&L 1998) \\
  thermal inertia $\Gamma$ & 15\,J\,m$^{-2}$\,s$^{-0.5}$\,K$^{-1}$  (M\"uller et al.\ \cite{mueller99})\\
  beaming parameter & $f=0.6$, $\rho=0.7$ (M\&L 2002) \\
   \noalign{\smallskip} \hline \hline \noalign{\smallskip}
   \multicolumn{2}{c}{TPM Predictions} \\
   \noalign{\smallskip} \hline \noalign{\smallskip}
  FD at 3180\,$\mu$m   & 40$\pm$4\,mJy ($\epsilon (\lambda)$, M\&L 1998)\\
  at observation epoch & 35$\pm$3\,mJy ($\epsilon (\lambda)$, but with \\
  (rotation averaged)  & $\epsilon \sim$ 0.7 at submm/mm)\\
   \noalign{\smallskip} \hline \noalign{\smallskip}
  mm-lc amplitude & $\pm$ 1\,mJy (= $\pm$ 2\%) \\
  TPM prediction  & (shape-dominated) \\
   \noalign{\smallskip} \hline \noalign{\smallskip}
  \end{tabular}
  \end{center}
\end{table}

\begin{figure}[h!tb]
    \begin{center}
    \resizebox{\hsize}{!}{\includegraphics*{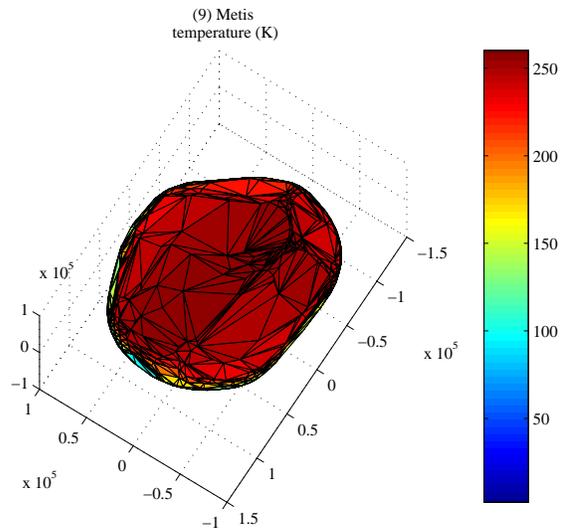}}
    \caption[]{Temperature distribution on the surface
               of (9)~Metis, based on the shape model (derived from lightcurve 
	       inversion techniques) in combination with
              the thermophysical parameters and calculated
	      for 2004-Oct-13 09:00:00 UT.}
    \label{fig:metis_tpm}
    \end{center}
\end{figure}

Storrs et al. (\cite{storrs05}) derived from WF/PC HST images a size of
$222\times182\pm12$\,km and an albedo of $p_\mathrm{V}=0.108\pm0.006$.
Tedesco et al. (\cite{tedesco02a}) calculated, via a simple radiometric method
and from a one-epoch MSX measurement in 4 bands between 4.2 and 26\,$\mu$m,
an effective diameter of $D_\mathrm{eff}=153.62\pm4.14$\,km and an albedo
of $p_\mathrm{H}=0.2307\pm0.0123$. But both results, the HST and the MSX,
are only related to an instantaneous cross section rather than
a reliable effective size.
Mitchell et al. (\cite{mitchell95}) combined radiometric, lightcurve and
occulation data, resulting in a model ellipsoid of $215\times
170\times135$\,km ($\pm$15\%). The corresponding effective diameter
is $D_{\mathrm{eff}} = 2(abc)^{1/3}= 170.2$\,km.
We re-determined the diameter and albedo values using the
Torppa et al. (\cite{torppa03}) shape and spin vector in the TPM code
together with the colour-corrected and calibrated
MSX flux densities (Tedesco et al. \cite{tedesco02a}), the ISO data point
(Lagerros et al. \cite{lagerros99}) and ground-based N-band
($\lambda = 7.5-14\,\mu m$) and Q-band ($\lambda = 16-28\,\mu m$)
observations by Hansen (\cite{hansen76}).
The weighted mean values from our radiometric
analysis are $D_{\mathrm{eff}} =171.9\pm13.0$\,km and
$p_\mathrm{V}=0.19\pm0.03$. The corresponding largest dimensions of the shape model
are $218\times180\times129$\,km, in excellent agreement
with the numbers by Mitchell et al. (\cite{mitchell95}). 

An independent confirmation of the albedo was published by
Nakayama et al. (\cite{nakayama00}) based on photo-polarimetric
observations over many phase angles. They derived an albedo
of 0.15. Adaptive optic images of (9)~Metis at 2 different epochs
(Marchis et al. \cite{marchis06})
prove the good quality of the shape model. The AO average 
size of 181\,km is also in agreement with our effective diameter
value. Nevertheless, the irregular shape and the surface variegations
are sources of uncertainties for the size as well as for the albedo
values resulting from different techniques.

For our subsequent calculations we used an effective diameter of 176$\pm$8\,km
and an albedo of 0.18$\pm$0.04. 
We applied the wavelength-dependent emissivity model for large
regolith-covered asteroids as determined
by M\&L (\cite{mueller98}). This emissivity model has values of 0.9
at mid-/far-IR wavelengths and 0.8 in the submm-/mm-range.
Figure~\ref{fig:metis_tpm} shows the temperature distribution picture
on top of the shape model for the epoch of the CA observations.

%\clearpage
%\newpage

%______________________________________________________________
\section{Discussion}
\label{sec:discussion}

\subsection{Asteroid (4)~Vesta}

\begin{figure}[h!tb]
    \begin{center}
    \rotatebox{90}{\resizebox{!}{\hsize}{\includegraphics{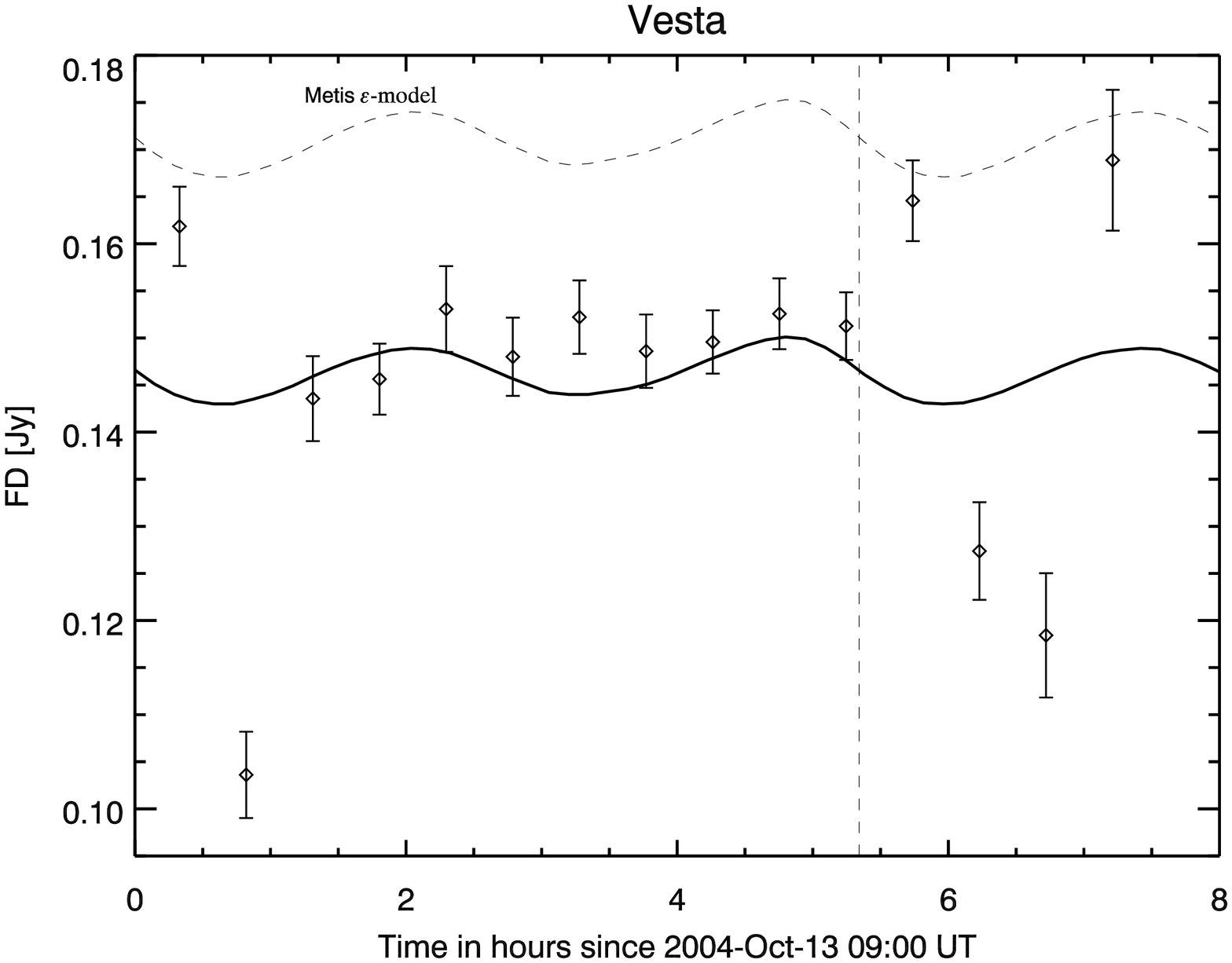}}}
    \rotatebox{90}{\resizebox{!}{\hsize}{\includegraphics{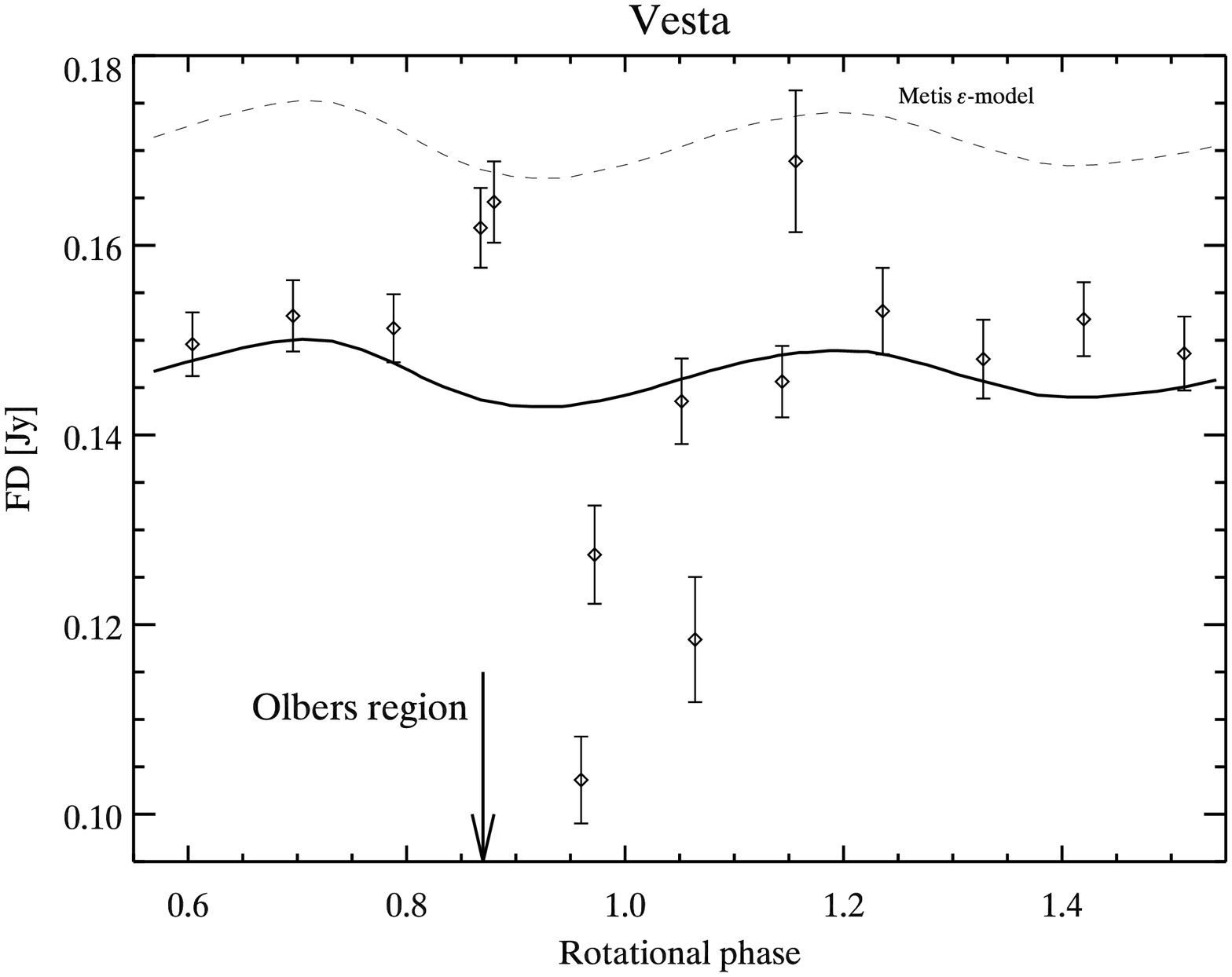}}}
    \caption[]{Top: Absolute flux prediction over time (solid line)
    together with the measurements (see Table~\ref{tbl:tbl3}).
    We used the lower emissivity of 0.6 at mm-wavelength
    together with a size of 535\,km and an albedo of 0.35.
    The vertical dashed line indicates one rotational period.
              Bottom: The same as in the top, but now plotted
	      against the rotational phase. For clarity reasons we
	      show the data at rotational phases between 0.55 and 1.55,
              instead of [0,1]. The zero point in rotational phase
              is connected to the definition in Table~\ref{tbl:tpm_vesta}.
	      For the dashed lines we used the emissivity model for (9)~Metis.
              The arrow refers to the rotational phase when the Olbers feature
              crosses the sub-Earth meridian.}
    \label{fig:vesta_obsmod}
    \end{center}
\end{figure}

Based on the input parameters in Table~\ref{tbl:tpm_vesta} we obtained a 3180\,$\mu$m flux density
of 165...175\,mJy, depending on the above albedo/diameter combinations.
Lowering the mm-emissivity also to a value of 0.6 would give flux densities
of 140-150\,mJy (averaged over one rotation period).
The albedo uncertainty does not affect the flux
prediction very much (10\% uncertainty produces a 1\,mJy flux change).
The diameter is more relevant: 2.5\% uncertainty would bring both
diameter solutions into agreement, but cause flux differences of
roughly 5\%, i.e., about 8\,mJy (see Table~\ref{tbl:tpm_vesta}).

The measured values favor therefore the solution with the 
lower emissivity at mm-wavelength, i.e., a constant low emissivity
of around 0.6 at wavelengths between the submm-range and a few mm.
In addition to our favoured solution (solid line in Fig.~\ref{fig:vesta_obsmod})
we also show the best emissivity model for (9)~Metis as a dashed line.

The TPM lightcurve amplitude calculation is shown in Fig.~\ref{fig:vesta_obsmod}
together with the observed data points.
This model lightcurve (solid line) is only reflecting the flux variation due to the
change in cross-section during the rotation of (4)~Vesta. There might be
an additional component due to albedo variations, but even very
strong variations could only produce very weak lightcurve features.
A simulated dramatic albedo drop on one hemisphere from 0.35 to 0.20 would cause
a flux change of about 3\,mJy, still within the shape-introduced effects.

A lack of insulating dust regolith on some parts of the surface
would have much more dramatic effects on the lightcurve: Under the given
observing and illumination geometry the change from a thick dust layer
with low thermal inertia to a rocky surface with high thermal inertia
could lower the flux by up to 20\,mJy. Such an effect might explain
the reproducible flux drop in Fig.~\ref{fig:vesta_obsmod} (bottom). On the other hand,
the thermal lightcurve at far-IR wavelength does not show any deviation
from the shape-introduced lightcurve of that kind (B.\ Schulz, priv.\
comm.). Redman et al. (\cite{redman92}) found that the 1\,mm light-curve
is apparently dominated by the triaxial shape, without any significant contributions
from the optical albedo spots. This excludes the theory of a pronounced
change in the surface texture. But the solution could be in a change
of the grain size distribution. Redman et al. (\cite{redman92}) showed
that scattering by gains within the regolith can reduce the emissivity
in a wavelength dependent fashion. This mechanism has also been proposed
to explain the low apparent emissivity of the Moon (Simpson et al. \cite{simpson81}).
The scattering becomes effective
at wavelength shorter than 2\,$\pi$\,$a$, with $a$ being the grain size.
Redman et al. (\cite{redman92}) 
speculated therefore that within (4)~Vesta's regolith there must be a large population
of particles around 100\,$\mu$m in size. We would follow their interpretation except
for the region on the surface which is responsible for the sudden flux drop
in the mm-light-curve. At that position there must be a grain population with
a predominantly larger particle size. Particles with sizes of several 100
$\mu$m would absorb more at mm-wavelengths without affecting very much the
emissivity in the far-IR or submm-range. We believe that the larger particles
might be related to a younger
surface, possibly due to a recent impact. These particles are not yet
processed by space weathering and impacts of micro-meteorites.

Barrera-Pineda et al.\ (\cite{barrera02}) also indicated that they saw
at 870\,$\mu$m a lightcurve with a 20\% amplitude and which varied
inversely to the visible lightcurves. Unfortunately, these observations
are not publicly available for independent investigations and to separate shape and emissivity influences.
From a comparision with the HST images of (4)~Vesta by Thomas et al. (\cite{thomas97})
they concluded that high temperatures are observed in regions of low albedo
and low temperatures in regions of high albedo. Furthermore, they saw a marked
difference in the thermal emission between the two "hemispheres" as outlined
in the geologic maps of Binzel et al. (\cite{binzel97}).
These considerations make the hint of a rising emissivity for (4)~Vesta
(Sect.~\ref{sec:obs_res}) all the more intriguing.

(4)~Vesta's measured fluxes at 93.0 and 95.5\,Ghz are the same (see
Table~\ref{tbl:tbl4}), and so somewhat at odds with the notion of a grey body
with a $S_I \propto \nu^2$ dependence as predicted by the TPM.
The TPM in fact predicts a roughly 8\,mJy drop in (4)~Vesta's flux from
95.5 to 93.0\,Ghz.  The red "colour" for (4)~Vesta in Table~\ref{tbl:tbl4} effectively
means that (4)~Vesta's low mm-emissivity might be rising again at longer
wavelengths.  In contrast, the frequency data for (9)~Metis are entirely
consistent with a grey body (constant emissivity) prediction from
the TPM.  If these 3\,mm colours can be confirmed, it would mean
that we could establish an upper
limit for the particle size in the scattering regolith.
What would be even more interesting is if we could see how the
amplitude of (4)~Vesta's lightcurve variations might change with
wavelength, allowing us to conduct this analysis over different
features of (4)~Vesta's surface.  However we emphasise that the
detection of a rising emissivity for (4)~Vesta towards longer
wavelengths is marginal (a 1.2-$\sigma$ result, or 3-$\sigma$
away from Olbers), and needs to be confirmed.

Zellner et al. (\cite{zellner97}) and Binzel et al. (\cite{binzel97})
reported on basis of HST images of (4)~Vesta a surface feature with
lower albedo. They called it "Olbers" and they speculated that
this might be related to (4)~Vesta's ancient basaltic crust. The
exact same rotational phase during our observations corresponds
to the two fluxes well above the predicted lightcurve
(at around rotatational phase 0.87 in Fig.~\ref{fig:vesta_obsmod} bottom).
The low albedo, probably in combination with slightly different
surface material properties, could explain the higher
mm-fluxes. The strong flux drop in the lightcurve in
Fig.~\ref{fig:vesta_obsmod} (bottom) occurs when the Olbers region
has moved to the side by about 40$^{\circ}$ in rotational phase, i.e.,
at 1.0 in Fig.~\ref{fig:vesta_obsmod} (bottom). It might well be connected to
the Olbers structure, e.g., ejecta material from an impact which
is deposited only on one side of a crater. We cannot confirm
a pronounced hemispherical difference as it was seen by Barrera-Pineda et
al.\ (\cite{barrera02}) from our data set. Instead, we attribute the
strong emissivity variations to the Olbers structure and a neighbouring region.
 
\subsection{Asteroid (9)~Metis}

\begin{figure}[h!tb]
    \begin{center}
    \rotatebox{90}{\resizebox{!}{\hsize}{\includegraphics{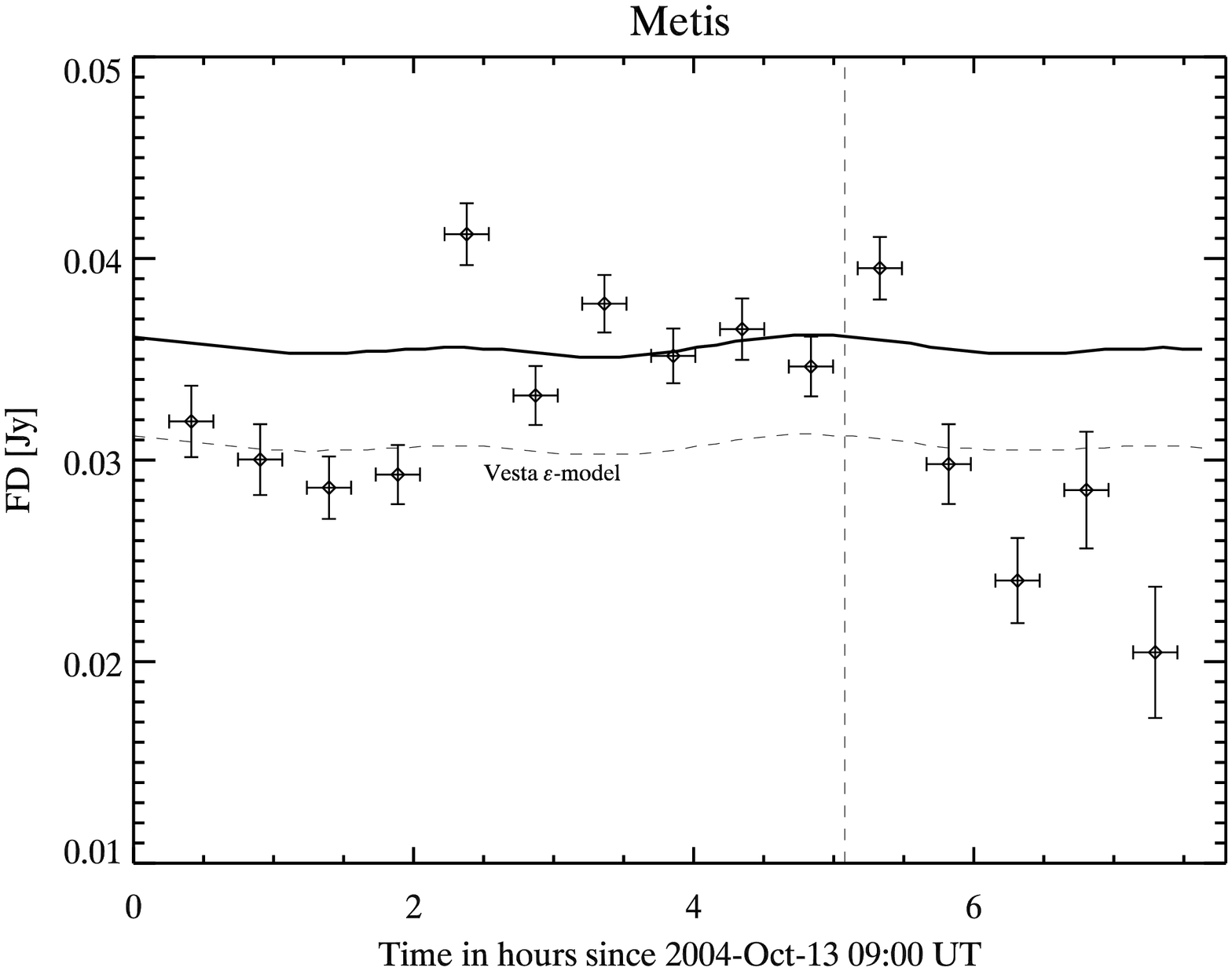}}}
    \rotatebox{90}{\resizebox{!}{\hsize}{\includegraphics{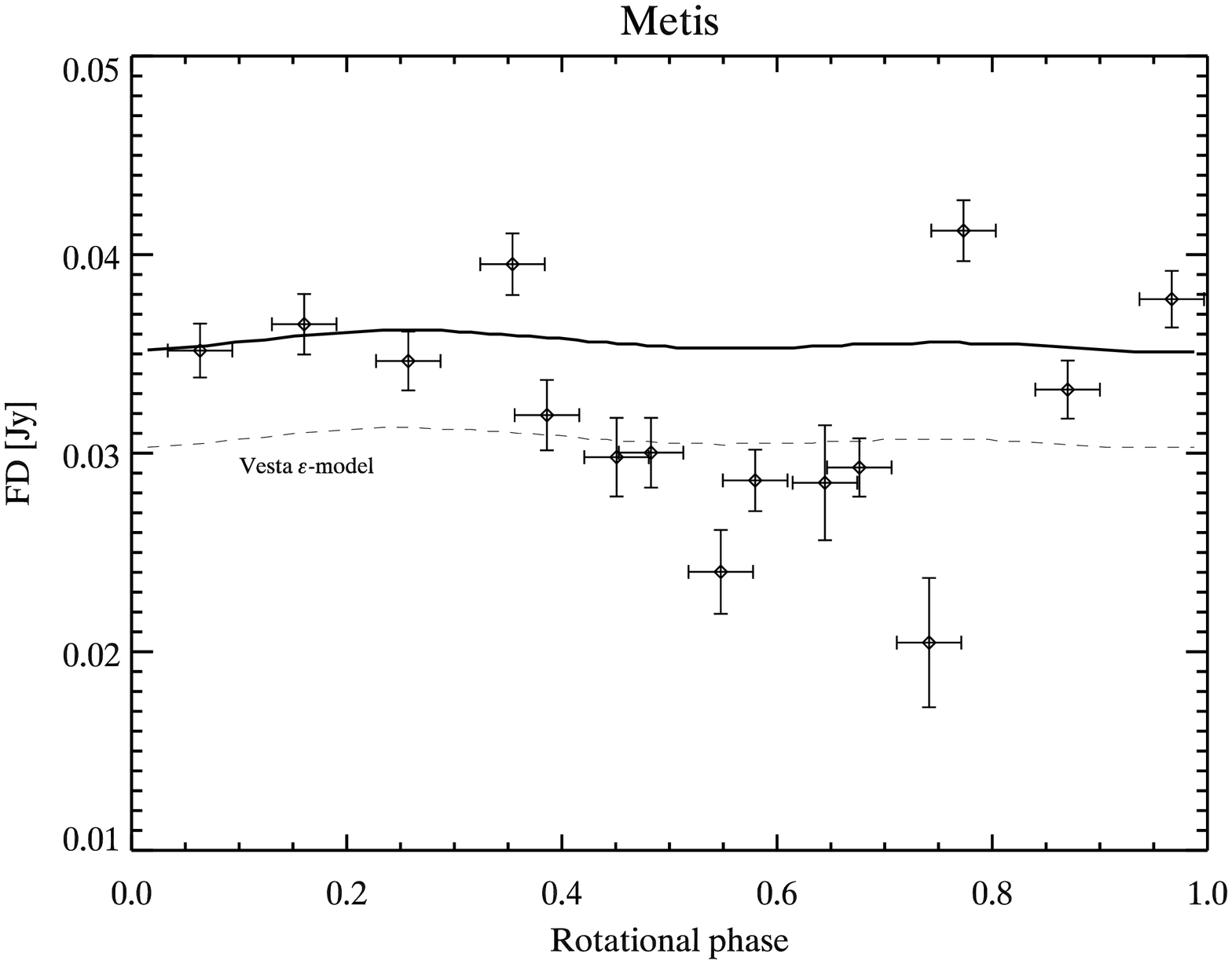}}}
    \caption[]{Top: Absolute flux prediction over time (solid line)
    together with the measurements (see Table~\ref{tbl:tbl3}).
    We used a wavelength-dependent emissivity model
    together with a size of 172\,km and an albedo of 0.16.
    Bottom: The same as in the top, but now plotted
	      against the rotational phase. The zero point in rotational phase
              is connected to the definition in Table~\ref{tbl:tpm_metis}.
              The dashed line indicates a model prediction with an emissivity
              of 0.6. This clearly shows the emissivity heterogeneity on the
              surface of (9)~Metis. For the dashed lines we used the emissivity
	      model for (4)~Vesta.}
    \label{fig:metis_obsmod}
    \end{center}
\end{figure}

As for (4)~Vesta, the albedo uncertainty does not affect the flux
prediction very much: the diameter is the dominant factor.
The model input parameters and some results are given in Table~\ref{tbl:tpm_metis}.
Based on these input parameters we obtained a 3180\,$\mu$m flux density
of 40$\pm$4\,mJy, depending on the above albedo/diameter combinations.
Lowering the mm-emissivity to a value of 0.7 would give flux densities
of 35$\pm$3\,mJy (averaged over one rotation period).
An emissivity model as used for (4)~Vesta produces rotational averaged
fluxes of about 30$\pm$3\,mJy.
The measured values favor therefore the solution with the 
lower emissivity, i.e., an emissivity of around 0.7 at mm-wavelengths.
In Fig.~\ref{fig:metis_obsmod} we show predictions for emissivities of
0.7 (solid line) and 0.6 (dashed line), which corresponds to (4)~Vesta's
emissivity model.

The predicted shape-introduced lightcurve variation under the given
illumination and observing geometry is about $\pm$2\%. Figure~\ref{fig:metis_obsmod}
shows the predicted flux variation for the observed period
in combination with the measurements from Table~\ref{tbl:tbl3}.

The measured lightcurve amplitude of about 24\% peak-to-peak
cannot be explained by the shape model.
The thermal inertia has almost no effect on the lightcurve
amplitude at these long wavelengths. Albedo spots can also
be excluded as a cause for the large lightcurve amplitudes since
the mm-emission is almost independent of albedo.
The only possible reason
seems to be that at certain rotational phases we see additional
emission coming from subsurface layers, meaning that the
regolith properties are inhomogeneous over the surface.
Similar to the explanations for (4)~Vesta, regions with young surface material
with predominantly larger particle sizes could lower the
emissivity significantly and produce very pronounced mm-lightcurve
effects which would be seen at visual wavelengths. In fact, studies
by Nakayama et al. (\cite{nakayama00}) and Storrs et al. (\cite{storrs99})
indicated a heterogeneous surface for (9)~Metis and the shape model (Torppa et al.
\cite{torppa03}) has some sharp features, but the shape model fits
nicely the available visual lightcurve without albedo variegations
on the surface.
Marchis et al. (\cite{marchis06}) reported a bright structure in their
Keck-AO images with a contrast of 20\% in a Kp broadband filter (1.95-2.30\,$\mu$m).
This prominent surface marking was visible both images which were separated by
about 2\,hours, taken at rotational phases of 0.42 and 0.81. These images were
taken only a few days after our CA observations under a similar aspect angle.
It might well be that this bright feature in reflectance has a significantly
lower mm-emissivity than the rest of the surface and therefore this structure
might cause the light-curve minimum in Fig.~\ref{fig:metis_obsmod} (bottom)
at rotational phases at around 0.55 $\pm$0.15. The high contrast dark
feature in the Keck image at rotational phase 0.81 might then be
responsible for the abrupt flux change at the end of the lightcurve
minimum in Fig.~\ref{fig:metis_obsmod} (bottom).

%______________________________________________________________
\section{Conclusion}

The thermophysical model has been very successful in fitting the observed spectral
energy distributions of main-belt asteroids across a wide range of wavelengths,
from $\sim$ 5\,$\mu$m to $\sim$ 1000\,$\mu$m, namely across the brightest
parts of their Planck functions (M\&L \cite{mueller98}; \cite{mueller02}).
At those wavelengths, the thermal behaviour
is dominated by shape, albedo and thermal properties of the surface regolith.
At millimetre-to-cm wavelengths albedo and thermal properties are less important and
instead the emissivity of the surface material plays an important role. Especially
the grain size distribution seems to lower the emissivity in the 90\,GHz range
and variations of the grain sizes apparently dominate the rotational flux changes.
This has been the first serious test of the model at much longer wavelengths than those for which it was
designed, and is therefore an opportunity not only to refine the model, but to learn more
about the surface properties of these asteroids.

The emissivity of (4)~Vesta and (9)~Metis at the observed frequency range
between 93 and 95.5\,GHz is significantly lower than at wavelength ranges
between the mid-IR and the sub-millimetre range.
This can be explained by regolith properties which allow us to see
colder layers below the surface. In other words, the grain sizes
are comparable to the observing wavelength and internal reflection and
scattering processes lower the emissivity in a wavelength-dependent
fashion. Our measurements also show for the first time that 
surface heterogeneities affect the 3\,mm-lightcurves dramatically.
At shorter wavelengths below 1\,mm, lightcurves are usually dominated
by the object's shape. The simple existence of a surface regolith
controls to a large extend the spectral energy distribution, but 
at 3\,mm differences in regolith properties can be revealed and
disentangled from the shape and albedo effects. Bright surface spots at
visual or near-IR wavelengths seem to correspond to low emissivity
parts in the 3\,mm lightcurves. The very dark Olbers region on
(4)~Vesta might be the cause of an emissivity increase above the
shape-introduced flux changes. Variations in emissivity at longer
wavelengths -- especially for (4)~Vesta -- need to be confirmed,
in order to model regolith properties more rigorously.
More observations of other targets, including also some lower albedo
objects, are needed to establish possible relations between the albedo
and the 3~millimetre emissivity. It would also be interesting to see
if apparently homogeneous objects follow the predicted lightcurves
or if they also show heterogeneities which are not present at
visual and near-IR wavelengths. But such observations require stable
conditions and careful calibration procedures during a significant part
of the asteroid's rotation period. Along with other mm-interferometers,
the ATCA has the potential to contribute significantly to these goals.

\begin{acknowledgements}

We would like to thank the ATNF TAC for the allocation of telescope time to an
unusual project, and to the staff of the Compact Array for their usual
outstanding help in meeting our goals. PJB gratefully acknowledges the
Institute for Astronomy in the School of Physics at Sydney University
for their support of the observations and analysis, and is also
delighted to thank Yasuo Fukui and his NANTEN group at Nagoya
University, for their hospitality during the conclusion of some of
this work.
We would also like to thank Mikko Kaasalainen for providing both shape models
in an easily usable format and for support in the shape model implementation.
\end{acknowledgements}

\end{document}